\documentclass[preprint, a4paper, times]{aastex61}
\usepackage{natbib}
\usepackage{amsmath}
\renewcommand\vec{\boldsymbol}
\renewcommand\tensor{\mathbf}

\begin{document}

\title{Interstellar neutral helium in the heliosphere from IBEX observations. V. Observations in IBEX-Lo ESA steps 1, 2, \& 3} 

\correspondingauthor{Maciej Bzowski}
\email{bzowski@cbk.waw.pl}

\author{Pawe{\l} Swaczyna}
\affiliation{Space Research Centre of the Polish Academy of Sciences (CBK PAN), Bartycka 18A, 00-716 Warsaw, Poland}

\author{Maciej Bzowski}
\affiliation{Space Research Centre of the Polish Academy of Sciences (CBK PAN), Bartycka 18A, 00-716 Warsaw, Poland}

\author{Marzena A. Kubiak}
\affiliation{Space Research Centre of the Polish Academy of Sciences (CBK PAN), Bartycka 18A, 00-716 Warsaw, Poland}

\author{Justyna M. Sok{\'o}{\l}}
\affiliation{Space Research Centre of the Polish Academy of Sciences (CBK PAN), Bartycka 18A, 00-716 Warsaw, Poland}

\author{Stephen A. Fuselier}
\affiliation{Southwest Research Institute, San Antonio, TX 78228, USA}
\affiliation{University of Texas at San Antonio, San Antonio, TX, USA}

\author{Andr{\'e} Galli}
\affiliation{Physics Institute, University of Bern, Bern, 3012, Switzerland}

\author{David Heirtzler}
\affiliation{University of New Hampshire, Durham, NH 03824, USA}

\author{Harald Kucharek}
\affiliation{University of New Hampshire, Durham, NH 03824, USA}

\author{David J. McComas}
\affiliation{Department of Astrophysical Sciences, Princeton University, Princeton, NJ 08544, USA}

\author{Eberhard M{\"o}bius}
\affiliation{University of New Hampshire, Durham, NH 03824, USA}

\author{Nathan A. Schwadron}
\affiliation{University of New Hampshire, Durham, NH 03824, USA}

\author{Peter Wurz}
\affiliation{Physics Institute, University of Bern, Bern, 3012, Switzerland}

\begin{abstract} 
Direct-sampling observations of interstellar neutral (ISN) He by Interstellar Boundary Explorer (IBEX) provide valuable insight into the physical state of and processes operating in the interstellar medium ahead of the heliosphere. The ISN He atom signals are observed at the four lowest ESA steps of the IBEX-Lo sensor. The observed signal is a mixture of the primary and secondary components of ISN He and H. Previously, only data from one of the ESA steps have been used. Here, we extended the analysis to data collected in the three lowest ESA steps with the strongest ISN He signal, for the observation seasons 2009--2015. The instrument sensitivity is modeled as a linear function of the atom impact speed onto the sensor's conversion surface separately for each ESA step of the instrument. We found that the sensitivity increases from lower to higher ESA steps, but within each of the ESA steps it is a decreasing function of the atom impact speed. This result may be influenced by the hydrogen contribution, which was not included in the adopted model, but seems to exist in the signal. We conclude that the currently accepted temperature of ISN He and velocity of the Sun through the interstellar medium do not need a revision, and we sketch a plan of further data analysis aiming at investigating ISN H and a better understanding of the population of ISN He originating in the outer heliosheath.
\end{abstract}


\section{Introduction} 
\label{sec:intro}

Direct sampling of interstellar neutral (ISN) gas in the heliosphere is a powerful tool for investigating the physical state of the interstellar matter ahead of the heliosphere \citep{mobius_etal:09a}. The ISN gas is a mixture of various elements, mostly hydrogen and helium \citep{frisch_etal:11a}. Strong ionization processes deplete the population of ISN H close to the Sun and thus ISN He is the most abundant species at 1 au \citep{rucinski_etal:03}. Due to the long mean free path of the neutral He atoms in the interstellar medium most of the observed ISN atoms are created in the unperturbed medium well ahead of the heliosphere \citep{bzowski_etal:17a}. These atoms are collectively called the primary ISN population. Moreover, an additional population of atoms is created in the interstellar medium modified due to interaction with the heliosphere --- the secondary ISN population \citep{baranov_malama:95}. 

Results of in-depth analyses of direct-sampling observations from GAS/Ulysses \citep{witte:04, bzowski_etal:14a, wood_etal:15a} and from IBEX-Lo \citep{bzowski_etal:15a, mccomas_etal:15a, mccomas_etal:15b, mobius_etal:15b, schwadron_etal:15a} provided the direction, speed, and temperature of the primary ISN He inflowing to the heliosphere. In addition to the primary ISN He, IBEX-Lo discovered the Warm Breeze \citep{bzowski_etal:12a, kubiak_etal:14a}. This most likely is the secondary population of ISN He, created via charge exchange between interstellar He$^+$ ions and ISN He atoms in the outer heliosheath \citep{kubiak_etal:16a, bzowski_etal:17a} and thus bears information on the plasma flow in the region beyond the heliopause. The signal measured by IBEX-Lo cannot be easily separated into the primary and secondary components. Moreover, due to the measurement technique used by IBEX-Lo \citep{mobius_etal:09a, park_etal:15a}, it is challenging to uniquely identify the counts registered by the IBEX-Lo instrument due to neutral He and H atoms. Consequently, the signal observed by IBEX-Lo is a sum of the primary and secondary populations of ISN He and ISN H. 

ISN H has been identified in IBEX-Lo observations \citep{saul_etal:12a, saul_etal:13a, schwadron_etal:13a}, but the understanding of the signal in different ESA steps of IBEX-Lo is still not satisfactory \citep{schwadron_etal:13a, katushkina_etal:15b}. One of the reasons was an inaccurate knowledge of the radiation pressure acting on H atoms in the heliosphere \citep[submitted]{kowalska-leszczynska_etal:17a}. The ability to resolve the ISN H and ISN He components in the observed signal is important because ISN H plays a vital role in general heliospheric studies and because neglecting the hydrogen contribution in the He signal may bias the inferred parameters of inflow of interstellar matter on the heliosphere, as illustrated by the differences in the estimates of these parameters by \citet{bzowski_etal:12a} and \citet{mobius_etal:12a} on one hand and by \citet{bzowski_etal:15a, leonard_etal:15a, mobius_etal:15b} and \citet{schwadron_etal:15a} on the other hand. A discussion of this topic was provided by \citet{swaczyna_etal:15a} and \citet{bzowski_etal:15a}.

IBEX-Lo is a time of flight mass spectrometer \citep{fuselier_etal:09b} that measures neutral atoms by registering their negative ions, created upon impact on a specially prepared conversion surface. The conversion surface is permanently covered with a thin layer of water, constantly replenished due to outgassing of material from the sensor and the spacecraft. Thus, in addition to negative ions from the direct ionization mechanism, negatively charged products of sputtering of material from the conversion surface are registered \citep{wurz_etal:06a}. In the case of H atoms, the first mechanism dominates. However, He atoms rarely form a negative ion \citep{wurz_etal:08a} and therefore they are detected owing to the second mechanism. The products sputtered by He atoms from the water layer make the IBEX-Lo instrument sensitive to He atoms with energies typical for ISN atoms at 1 au from the Sun. The ions sputtered by He atoms from this water layer are primarily H$^-$ and O$^-$ ions. Thus, the signal interpreted as coming from He atoms can originate both from H$^-$ ions sputtered off the conversion surface by the incoming He atoms and from real H atoms reflected and ionized on the conversion surface. Disentangling the two contributions requires a detailed analysis \citep{park_etal:16a}. Calibration of the sensitivity of IBEX-Lo to H atoms with various energies was carried out in the laboratory, but this was possible only within certain limitations due to challenges in obtaining a monoenergetic beam of neutral atoms with relevant energies \citep{wieser_wurz:05a} and because the water layer on the conversion surface may be different between the laboratory and space conditions. 

IBEX-Lo does not directly measure the energy of the impacting atoms. It detects ions created at the conversion surface with energies within the energy range set by the electrostatic analyzer (ESA) and rejects those with the energies outside this preselected range. Ions with different energies are observed when the instrument is switched to a different energy setting. The atoms directly converted to negative ions preserve most of their energies, but the ions sputtered from the conversion surface have a broad energy distribution, limited by the energy of the impacting atoms. As discussed later in the paper, ISN He is visible in the four lowest energy settings of the ESA (hereafter ESA steps), but so far only measurements from ESA step 2 have been used in the analyses of ISN He. In this paper, we extend the analysis to ESA steps 1 and 3. ESA step 4 is left out because of significantly lower count rates (by a factor of $\sim$5). A better insight into the physics of interaction of the heliosphere with interstellar matter could potentially be obtained if data from the three ESA steps with the highest count rates are used. The factor that has been preventing the use of data from all of them is the lack of relative calibration of the ESA steps for He atoms.

In the preliminary analyses \citep{mobius_etal:09a, bzowski_etal:12a, mobius_etal:12a} it was assumed that the sensitivity function to various energies in ESA step 2 is flat, i.e., that the sensitivity is independent of energy. However, \citet{kubiak_etal:14a, sokol_etal:15a}, and \citet{galli_etal:15a} showed evidence that most likely, there is a threshold for the sensitivity to He atoms at an energy between 19 and 38~eV. This is low compared to the $\sim$130 eV incident energy of the ISN He atoms. However, this threshold was shown to be particularly important for the analysis of the Warm Breeze \citep[][Figure 8]{kubiak_etal:14a} and for the attempt to find ISN He in observations carried out during the Fall ISN observation seasons \citep{galli_etal:15a}. 

In this paper we make the first attempt to simultaneously determine the inflow velocity vector, the temperature of ISN He, and the relative energy sensitivity characteristics of the IBEX-Lo detector to He atoms using data from the three lowest ESA steps. We use measurements from the ISN observation campaigns 2009 through 2015, i.e., one season more than in the previous analyses \citep{bzowski_etal:15a,mobius_etal:15b}. First, we discuss the data, the adopted models, and the analysis method (Section~\ref{sec:data_models}). Then, we present results of the analysis, including the relative energy sensitivity characteristics of the three ESA steps of IBEX-Lo (Section~\ref{sec:results}). Finally, we discuss some implications of the findings (Section~\ref{sec:discussion}). 

\section{Methods}
\label{sec:data_models}

\subsection{IBEX-Lo observations}
\label{sec:data}

\begin{figure} 
%\epsscale{1.15} 
\epsscale{.6} 
\plotone{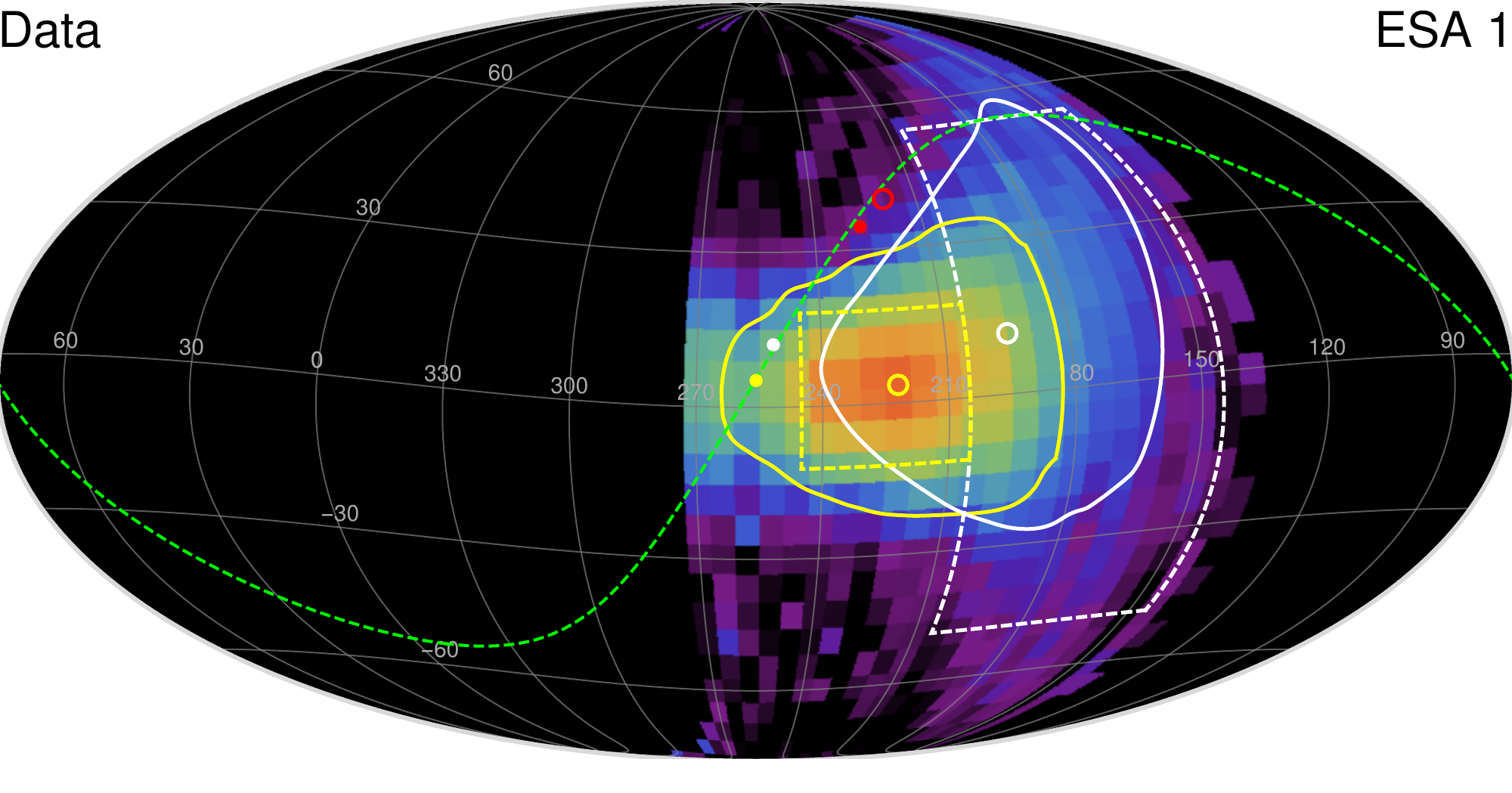} 
\plotone{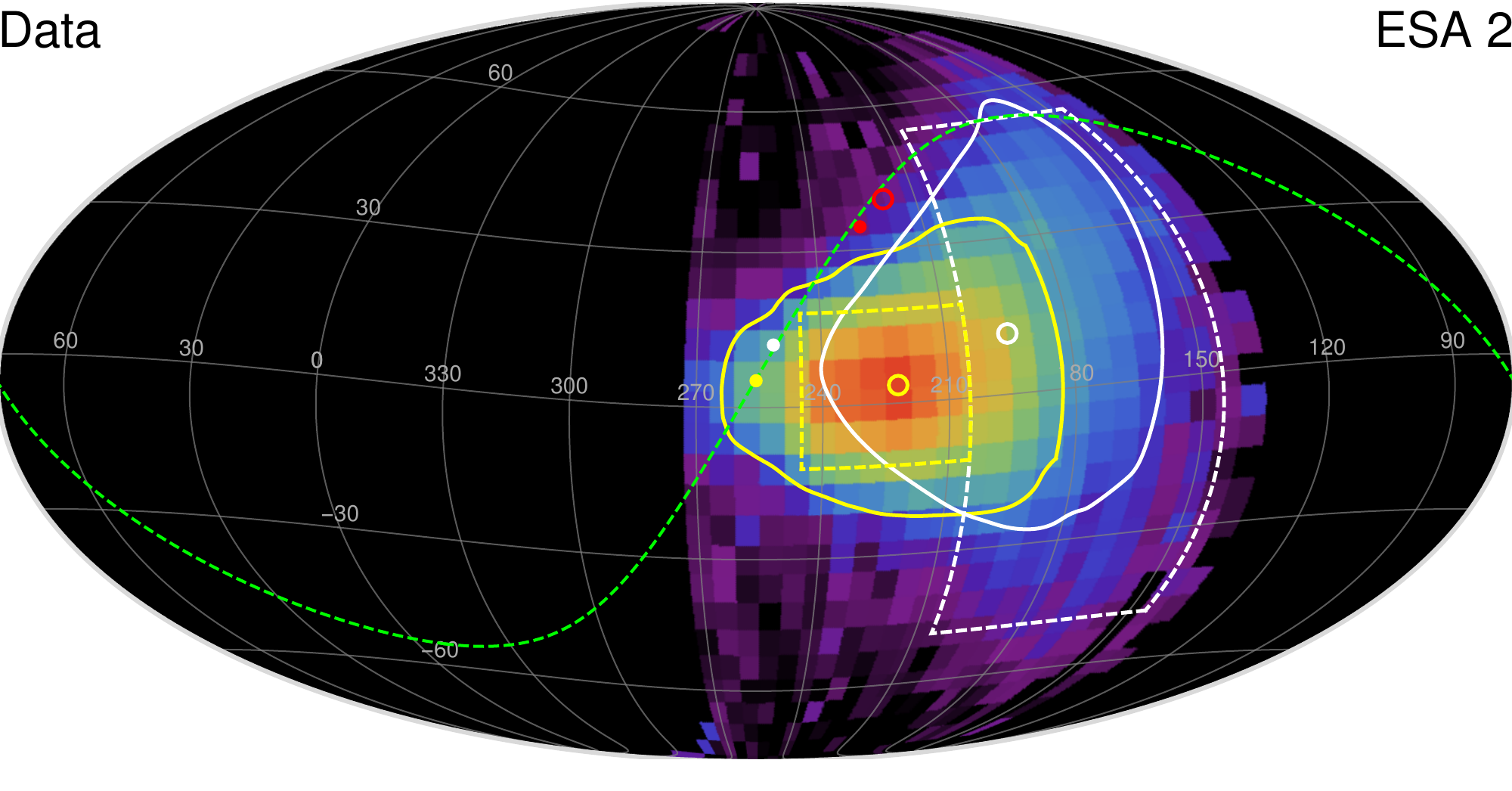}
\plotone{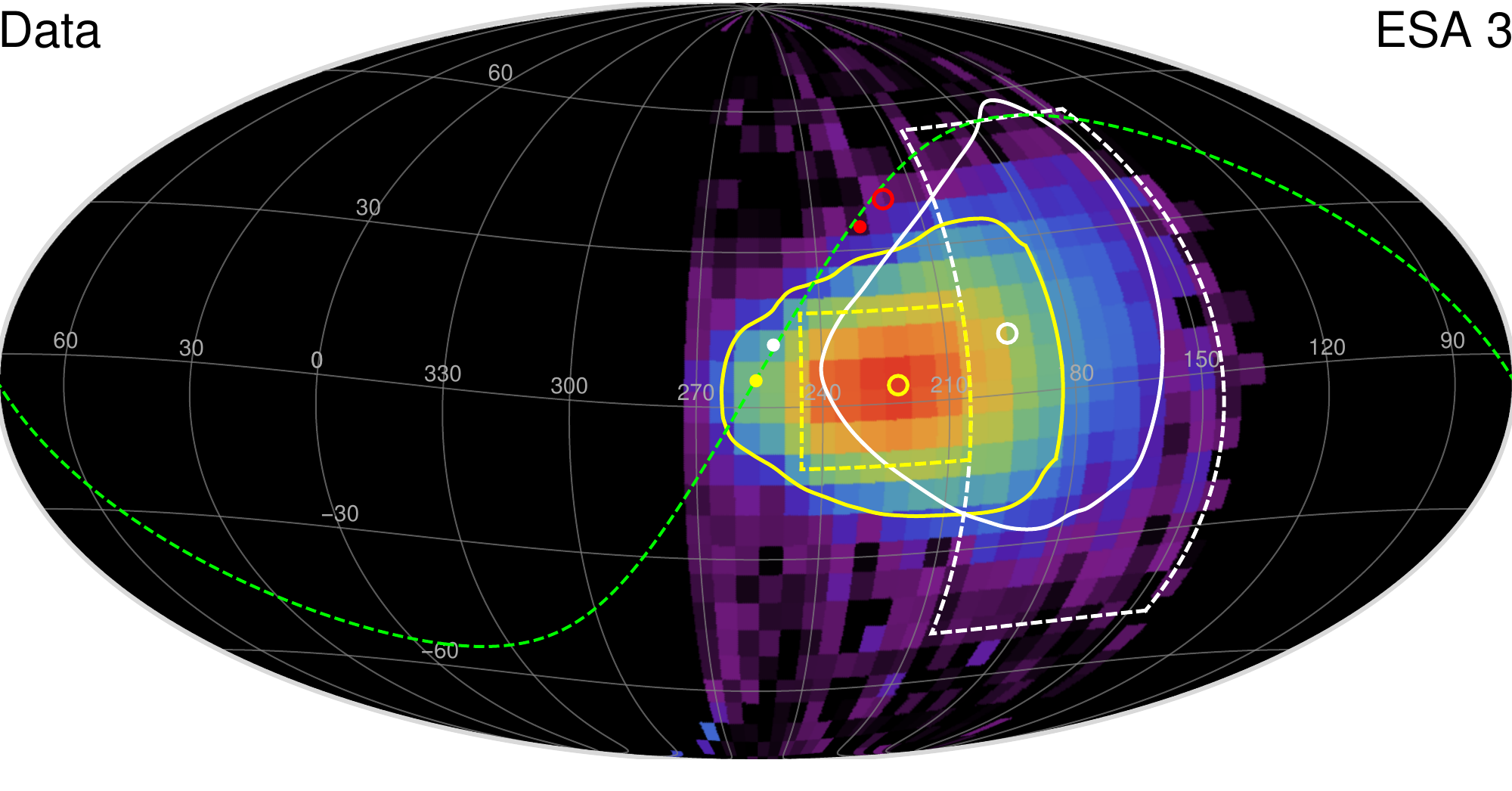}
\plotone{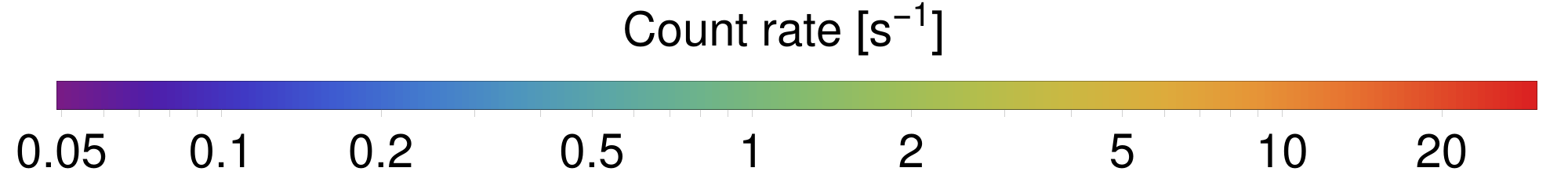}
\caption{Sky maps of the count rate in s$^{-1}$ due to ISN atoms observed by IBEX-Lo in ESA steps 1 (top), 2 (middle), and 3 (bottom) in the J2000 ecliptic coordinates centered at the nose direction in the IBEX-inertial frame, averaged over the ISN campaigns 2009 through 2015 (the sensitivity decrease after 2012 is compensated for, see text). The yellow and white dashed rectangles on the celestial sphere mark the data range used in the fitting of the primary ISN He in \citet{bzowski_etal:15a} as well as in the present paper, and in the fitting of the Warm Breeze in \citet{kubiak_etal:16a}, respectively. The solid contours mark where the contribution of the primary ISN He (yellow) and the Warm Breeze (white) signal in ESA step 2 is $\approx$0.1~s$^{-1}$; note that these contours are identical in all three panels to facilitate viewing the differences between the flux observed in the three ESA steps. The maxima of the two components (marked with open yellow and white circles) fall inside their respective contours. The dashed green line is the deflection plane of the secondary population as obtained by \citet{kubiak_etal:16a}. The unperturbed inflow directions of these two populations are shown as the solid yellow and white dots on this plane. The red circle represents the center of the IBEX ribbon \citep{funsten_etal:13a} and the red dot is the direction of the unperturbed interstellar magnetic field, derived by \citep{zirnstein_etal:16b} from global models of the heliosphere using the geometry of the ribbon as constraints.
}
\label{fig:map`E123}
\end{figure}

\begin{figure}
%\epsscale{1.15} 
\epsscale{.7} 
\plotone{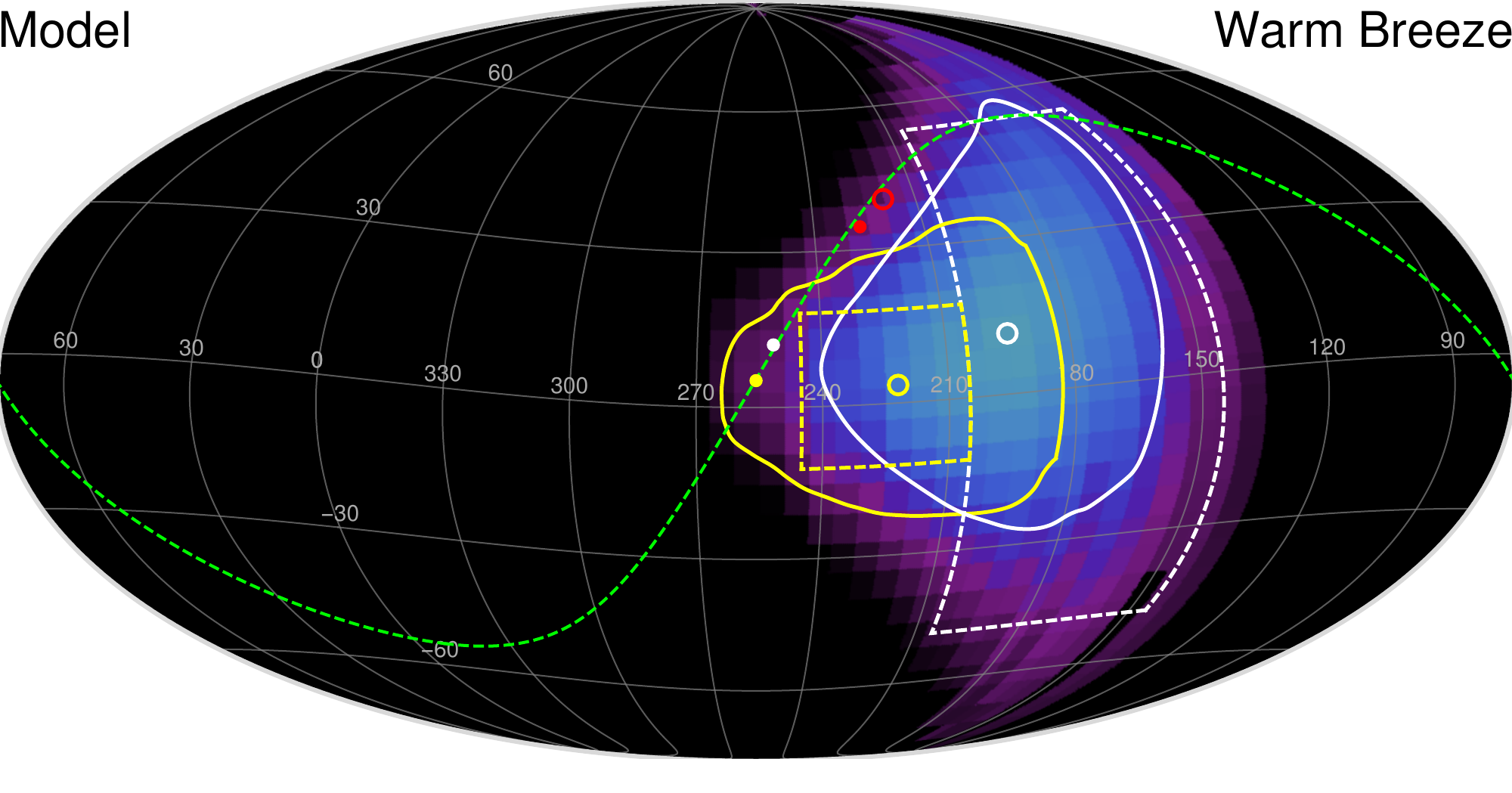} 
\plotone{legend_rate.pdf}
\caption{Simulated count rate due to the Warm Breeze in ESA step 2, calculated using the Warm Breeze inflow parameters from \citet{kubiak_etal:16a}, averaged over all ISN observation campaigns used in the present analysis. The lines, points, and contours are identical to those in Figure~\ref{fig:map`E123}. }
\label{fig:WBModel}
\end{figure}

IBEX-Lo observations are carried out in eight partially overlapping, logarithmically-spaced, sequentially switched energy channels (the ESA steps). The full width at half maximum (FWHM) of the ESA steps is $\Delta E \simeq 0.7 E$, where $E$ is the central energy of a given ESA step. For our purposes, the most relevant are the three lowest channels, i.e., ESA steps 1--3, with central energies 15, 29, and 55~eV, respectively \citep{fuselier_etal:09b, fuselier_etal:12a}. The absolute sensitivity of the instrument depends on the post-acceleration (PAC) voltage in the electrostatic analyzer. This value is generally kept constant, but once during the mission, after the fourth ISN campaign, it was reduced. As discussed by \citet{bzowski_etal:15a}, this resulted in an approximately two-fold reduction in the absolute sensitivity to He atoms. The central energies and widths of the ESA steps correspond to energies of incident H atoms that are converted to negative ions. The energy loss is small in this situation. Helium atoms are observed due to the sputtering of material from the conversion surface. This process results in a significantly larger energy difference between the incident He atom and the sputtered H$^-$ ion. The sputtered ions have a wide energy spectrum and are observed in the ESA steps with central energies much below the energies of the incident ISN He atoms.

Details of the operation of the IBEX-Lo instrument during the yearly campaigns of observations of ISN gas have been extensively presented in the literature \citep[e.g.,][]{mobius_etal:15a, bzowski_etal:15a} and will not be repeated here. In brief, IBEX is a spin-stabilized Earth satellite in a very elongated orbit \citep{mccomas_etal:09a}. To maintain the spin axis within a few degrees from the Sun, it was adjusted once per orbit during the first three years of IBEX operation, and is adjusted twice per orbit after the IBEX orbit change in 2011 \citep{mccomas_etal:11a}. The boresight of IBEX-Lo points perpendicular to the spin axis. For a given orientation of the spin axis, data are collected from the great circle of the sky visible to the instrument for several days when the spacecraft is sufficiently high above the magnetopause. The registered counts are binned into 6$\degr$ spin-angle bins. 

On the ground, the data are filtered against all known perturbations. Due to this filtering, some intervals of the observations are excluded. The intervals used in the analysis are referred to as ISN good times. In this paper, the previously used ISN good times \citep{leonard_etal:15a, mobius_etal:15a} are further restricted by adopting the intervals included in the good times obtained from an in-depth investigation of the sources of background and foreground, presented by \citet{galli_etal:15a, galli_etal:16a} and \citet{galli_etal:17a}. This restriction results in shorter accumulation times in several orbits and rejection of all data from orbit 237b. The list of orbits selected for this analysis is presented in Table~\ref{tab:orbits}. The background level was also revised following results from \citet{galli_etal:16a, galli_etal:17a}. The data subset used now is cleaner, which results in a slightly lower $\chi^2$ obtained in the present fitting than that by \citet{bzowski_etal:15a}, as will be evident from the results presented below.

\begin{deluxetable}{ll}
	\tablecaption{\label{tab:orbits}List of orbits selected for this analysis}
	\tablewidth{0pt}
	\tablehead{	\colhead{Year} & \colhead{Orbits} }
	\startdata
	2009 & 14, 15, 16, 17, 18, 19\\
	2010 & 63, 64, 65, 66, 67\\
	2011 & 110\tablenotemark{*}, 112\tablenotemark{*}, 113\tablenotemark{*}, 114\tablenotemark{*}, 115\\
	2012 & 153b\tablenotemark{*}, 154a\tablenotemark{*}, 154b\tablenotemark{*}, 156a\tablenotemark{*}, 156b, 157a, 157b, 158a\\
	2013 & 193a, 193b, 194a, 194b, 195a, 195b, 196a, 196b, 197a, 197b\\
	2014 & 233b, 234a, 234b, 235a, 235b, 236a, 236b, 237a, 238a\\
	2015 & 274a, 274b, 275a, 275b, 276a, 276b, 277b\\
	\enddata
	\tablenotetext{*}{data only from ESA step 2 (ESA step 1 \& 3 not observed due to high resolution mode) }
\end{deluxetable}

The data used in this analysis includes (1) intervals of the good times, and (2) the  numbers of counts due to ISN atoms, collected during these time intervals in each individual 6$\degr$ spin angle bin. From this information, we calculate mean counting rates for each bin for a given orbit. 

An extensive presentation of the data used for the ISN He analysis is available in \citet{bzowski_etal:15a}, and some important aspects of the data processing carried out before these data are used in the analysis were discussed by \citet{mobius_etal:15a, mobius_etal:15b}, and \citet{swaczyna_etal:15a}. The basis for fitting the parameters are orbit-averaged counting rates as a function of spin angle bins. For illustration purposes, however, in Figure~\ref{fig:map`E123} we show them as sky maps in the spacecraft inertial frame in ecliptic coordinates centered at the inflow direction $(255\fdg7, 5\fdg1)$, averaged over all observation seasons 2009 through 2015. This centering allows for direct comparison of the presented maps with the other maps of the atom fluxes observed by IBEX \citep[e.g.,][]{mccomas_etal:17a}. The count rate in each pixel is a weighted average of the $6\degr$ spin angle bins that fall into the pixel with the weights resulting from the Poisson uncertainty. The reduced instrument sensitivity after 2012 is compensated by factors of 2.2, 2.2, and 2.1 for ESA steps 1, 2, and 3, respectively (see Section~\ref{sec:results}). The maps are presented separately for the three ESA steps. In these maps, we mark the data range used for the present analysis of ISN He, as well as the range used by \citet{kubiak_etal:16a} in their analysis of the Warm Breeze component. The data used in the Warm Breeze fitting were taken between mid-November of the preceding year and mid-January. The data collected afterwards until the end of February are used in the ISN He analysis. The range of the data presented in the maps cover only the ISN seasons and thus spread only over part of the sky. The black region (longitudes outside the range $\sim$(145$\degr$, 270$\degr$) occupied by the ISN emission) are observed by IBEX-Lo but the data are not presented here because they only have contributions from inner heliosheath ENAs \citep{fuselier_etal:12a, schwadron_etal:14b, galli_etal:17a}.

It is important to realize that solar gravity significantly modifies the trajectories of ISN He atoms. In addition, IBEX is moving relative to the Sun and the ISN atoms it observes, thereby requiring a frame transformation for the observed velocities. Therefore, the direction of entry of an atom into the instrument collimator significantly differs from the direction of entry of the atom into the heliosphere. The mean magnitudes of this angular shift are different for the primary ISN He and the Warm Breeze populations, as illustrated in Figure~\ref{fig:map`E123}. The angle between the peak flux observed by IBEX and the inflow direction at the heliopause is $\sim$30$\degr$ for the primary ISN population and $\sim$54$\degr$ for the Warm Breeze population, which is indicated by the correspondingly colored circles and dots in this figure.

The observed ISN He flux distribution in the sky, as represented by the measured counting rate, is a result of a convolution of the angular distribution of the atoms entering the instrument with the collimator transmission function of IBEX-Lo \citep[Equation~32 in ][]{sokol_etal:15b}. Differences between maps obtained from individual observation campaigns result from several factors. (1) The ionization rate, and thus ionization losses, vary with solar activity. (2) The observation geometry varies from season to season (the spin axis orientations do not precisely repeat from one year to another, and the distribution of good times for equivalent orbits varies between observation seasons). (3) The IBEX motion around the Earth varies slightly for equivalent orbits, and additionally there was an important change of the IBEX orbit in 2011 \citep{mccomas_etal:11a}. All those elements are taken into account in the modeling of the IBEX signal. 

Inspection of the three panels of Figure~\ref{fig:map`E123} shows that, generally, the observed signal is comparable in shape and magnitude in all three ESA steps, but there is a clear, systematic reduction of the area in the sky occupied by the signal from lower to higher ESA steps. This effect is especially pronounced in the region occupied solely by the Warm Breeze, i.e., to the right of the ISN signal peak. To the left of the region used for analysis of ISN He, there is a region where a relatively strong signal is visible in the lowest ESA step, while the intensity decreases in ESA steps 2 and 3. This region is believed to be occupied mostly by ISN H and, as shown in the figure, has not been used in the analysis \citep{saul_etal:13a}. How much ISN H contributes to the signal has not been fully determined to date and is a subject of ongoing research.

The total distribution of ISN H is strongly modified in the outer heliosheath due to charge exchange between ISN H and the local disturbed plasma, so the temperatures and bulk velocity vectors of ISN H are different from those of ISN He already at the entrance to the heliosphere \citep[for in-depth discussion of these effects, see ][]{katushkina_etal:15a}. Inside the heliosphere, hydrogen atoms follow complex trajectories due to the action of solar radiation pressure. Radiation pressure is a function of the radial velocity of the atom due to the Doppler effect and of the magnitude of the solar Lyman-$\alpha$ flux, which varies over the solar cycle \citep{tarnopolski_bzowski:09}. Hydrogen atoms are subject to much stronger ionization losses than helium, and the ionization rate varies differently with time and heliolatitude than that for He. All these effects result in a different location of the peaks of the counting rates due to ISN H and ISN He in IBEX-Lo sky maps, and in different ratios of the H/He fluxes in different yearly observation campaigns. In particular, during some seasons (especially those close to the maximum of solar activity) ISN H may be practically unobservable for IBEX-Lo \citep{saul_etal:13a, galli_etal:17a}. 

Before fitting the parameters of the ISN He gas, it was necessary to subtract the Warm Breeze contribution from the signal. Even though this contribution is small within the data region adopted for analysis its presence would significantly affect the accuracy of the derivation of the ISN He inflow parameters \citep{swaczyna_etal:15a}. To that end, we calculated the expected count rate due to the Warm Breeze using the model parameters obtained by \citet{kubiak_etal:16a}, separately for each of the ISN observation campaigns. The time-averaged contribution from the Warm Breeze is shown in Figure~\ref{fig:WBModel}. To account for the differences between the sensitivities in the ESA steps used, the model count rate calculated for ESA step 2 was rescaled to steps 1 and 3 by factors $0.8 \pm 0.2$ and $1.0 \pm 0.3$, respectively. These factors were assessed based on the comparison of the observed Warm Breeze signals in the respective ESA steps. In this analysis we do not account for the energy dependent response in the Warm Breeze. This variation may be even more important for this population than for ISN He but we left it for future studies. Fortunately, the speed range of Warm Breeze atoms in the range used in the spin angle range used in this analysis is small (see Figure \ref{fig:map`speeds}). Because of the fit uncertainty of the Warm Breeze parameters, subtracting this model from the data modifies the uncertainty system of the data. This additional uncertainty was accounted for in the parameter fitting.

\subsection{Accounting for the energy characteristics of IBEX-Lo}
\label{sec:energyCharacteristics}

\begin{figure}
%\epsscale{1.15} 
\epsscale{.7} 
\plotone{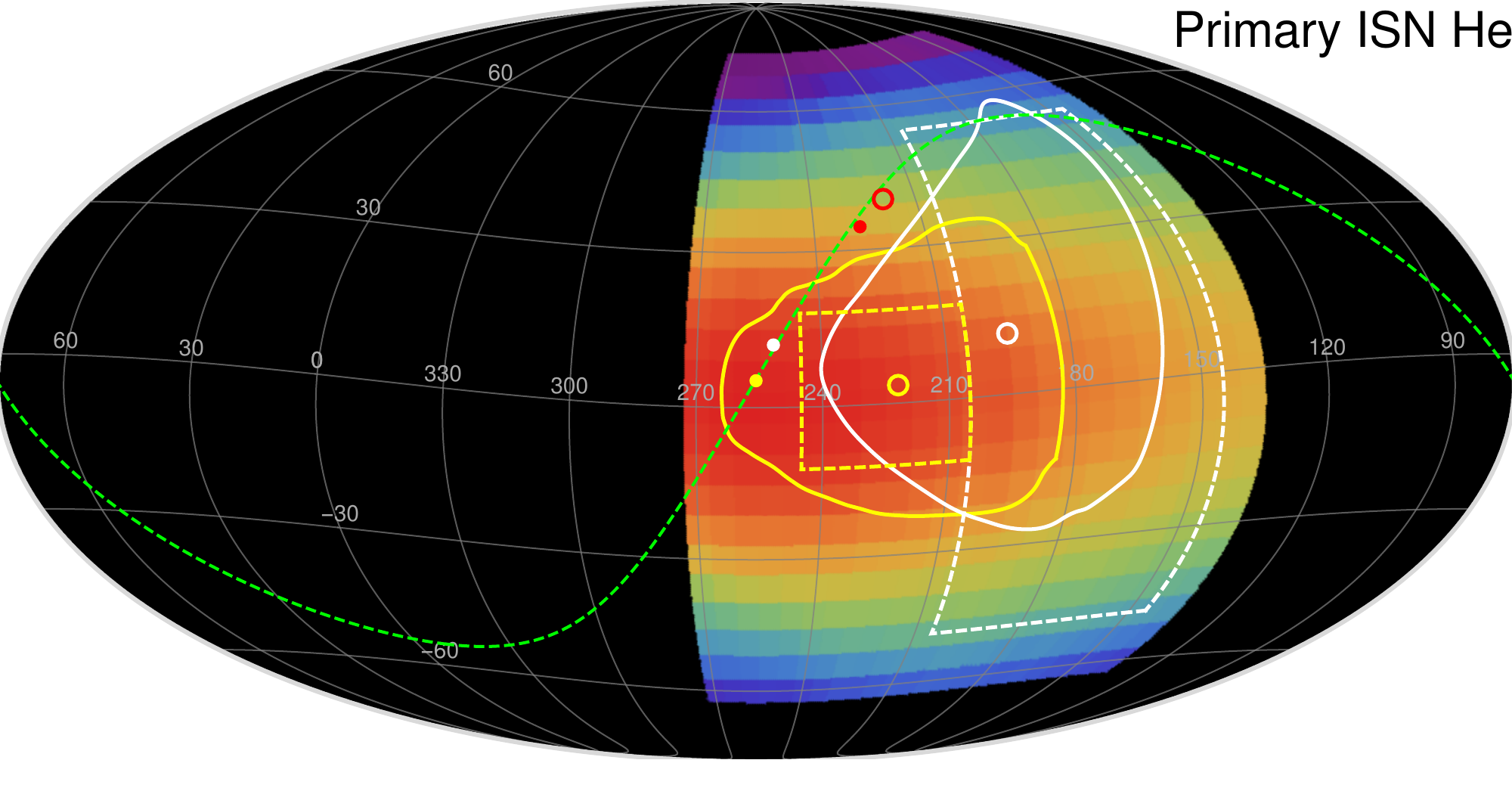} 
\plotone{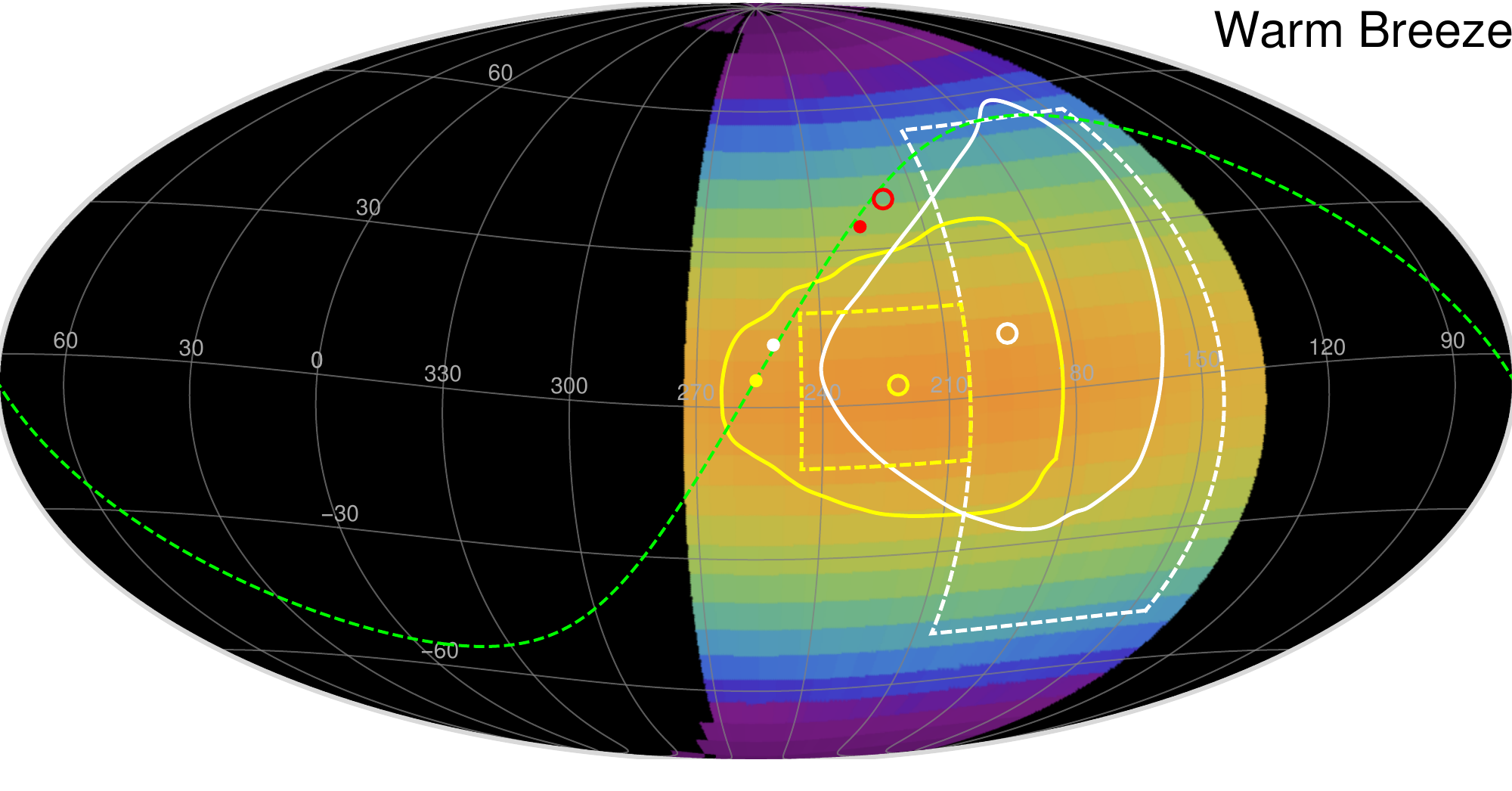} 
\plotone{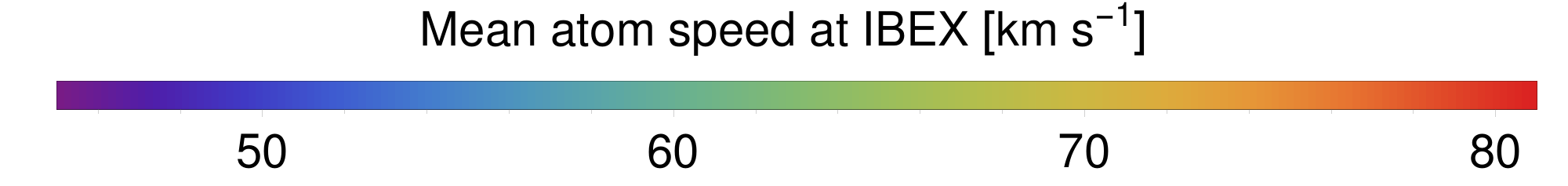} 
\caption{Sky maps of the mean speed of He atoms from the primary population (upper panel) and the Warm Breeze (lower panel) relative to IBEX. Different ecliptic latitudes correspond to different spin angle bins. The absolute speed scale in km~s$^{-1}$ is shown with the color bar. 
}
\label{fig:map`speeds}
\end{figure}

Fitting the data and inferring the flow parameters of ISN He is performed using forward modeling. In this approach, a model of the observed gas is calculated for an assumed set of ISN and instrument parameters, and the results are compared with the data. This modeling is done using the Warsaw Test Particle Model \citep[WTPM,][]{sokol_etal:15b}, which assumes a spatially homogeneous flow of Maxwell-Boltzmann gas everywhere beyond the heliopause. Inside the heliopause, the atoms follow heliocentric hyperbolic orbits and are subject to time-varying ionization losses due to solar photoionization and solar wind particles (electrons and protons), with the rates taken from \citet{bzowski_etal:13a, sokol_bzowski:14a}; and \citet{sokol_etal:16a}. 

The model used in the fitting adopts certain assumptions on the efficiency of converting the flux of atoms entering the instrument into a counting rate returned by this instrument. In the studies of the primary population of ISN He by \citet{bzowski_etal:15a} it was assumed that within ESA step 2 there is no energy dependence of the instrument sensitivity, but that the overall sensitivity could change from one observation season to another. The sensitivity level for each season was a free normalization parameter of the fit, further on referred to as the Y-norm (for ``yearly norm''). Based on the fit results it was demonstrated that the yearly sensitivities agreed with each other within the time intervals of identical PAC voltage in the instrument. Statistically, this result could be interpreted as no change in the sensitivity between individual seasons other than an approximately twofold reduction in the sensitivity due to the reduction of the PAC voltage in 2012. 

Based on the conclusion of the constancy of Y-norms within the intervals of identical PAC values, two normalization parameters for the two observation intervals with the two PAC values were adopted in the analysis of the Warm Breeze by \citet{kubiak_etal:16a}. Norms of this kind will be further on referred to as P-norms (for ``PAC Voltage norms''). 

In the present analysis, we used data from three energy steps instead of one and therefore we had to modify the approach to the energy sensitivity in the modeling. In reality, the energy response function for a given species depends on several factors: (1) sputtering efficiency, (2) energy distribution of the sputter products, (3) transmission efficiency for the sputter products through the ESA, (4) detection efficiency in the time-of-flight section of the detector. Some of these aspects are not fully understood for the actual in-flight environment. Factors (1) and (2) are independent of setting the instrument to a given ESA step because observations in different ESA steps are carried out in close succession, and the flux and speed distribution of the incoming atoms do not change over short times. As a result, the signal observed in ESA steps 1, 2, and 3 is created by the same flux of He atoms. Consequently, the mean counting rates obtained in a given spin angle bin in a given IBEX orbit in these ESA steps differ solely due to the different sensitivities of the instrument in these ESA steps (excluding the inevitable statistical scatter).  

From modeling, We found the range of the mean impact velocities of the observed ISN He atoms in each spin angle bin. The resulting mean impact velocities are presented in Figure~\ref{fig:map`speeds}. In this figure, we show that each spin angle bin has its own quite narrow range of atom speeds. Within the range of the data used for the ISN and Warm Breeze analyses, the mean impact velocities for the ISN He atoms range from 78 to 81~$\text{km s}^{-1}$, and for the Warm Breeze from 60 to 75~$\text{km s}^{-1}$. Therefore this unknown sensitivity function can be approximated by a linear function of speed over a relatively narrow speed range. The parameters of these linear relations, different for each ESA step and for each PAC voltage, will be free parameters in the fits. Within the data range used, the atom energies are $\sim$127--137 eV, well above the threshold energy, discussed in the Introduction, and thus the issue of the magnitude of this threshold speed does not affect our analysis. 

\subsection{Scaling functions}
\label{sec:model}

The model we adopted in the simulations is an extension of the version of the Warsaw Test Particle Model (WTPM) discussed in detail by \citet{sokol_etal:15b}. The calculation proceeds identically as discussed in that paper up to Equation~(44), which defines the particle scalar flux $F_i(\vec{\pi})$ in a given IBEX orbit, averaged over the width of the spin angle bin and good time intervals. Index $i$ corresponds to the index of a given data point, and $\vec{\pi}$ is the ISN He parameter set chosen to calculate the given model (the inflow velocity vector and the temperature). Alongside this calculation, we now calculate the mean impact speed for He atoms for each bin $v_i(\vec{\pi})$. The difference in the calculation of the count rate that we introduce in comparison with \citet{sokol_etal:15b} and \citet{swaczyna_etal:15a} is the more general scaling function $S(i, v, \vec{q})$, used here instead of the constant scaling factor adopted previously. The scaling function (expressed in cm$^2$sr) is a function of the mean atom impact velocity and converts the physical flux $F_i$ (in $\text{cm}^{-2}\text{s}^{-1}\text{sr}^{-1}$) to the simulated count rate $g_i$ (in s$^{-1}$): 
\begin{equation}
g_{i}(\vec{\pi},\vec{q}) = S(i, v_i(\vec{\pi}), \vec{q}) F_i(\vec{\pi}),
\label{eq:defCntRate}
\end{equation}
where $v_i$ is the characteristic speed for a given orbit and spin angle bin, which is now calculated alongside $F_i$, and $\vec{q}$ is a vector of parameters that define the scaling function. Importantly, $F_i$ is a physical characteristic of the atoms observed in this spin angle bin and orbit, which of course does not depend on the ESA step of the instrument. In addition to $F_i$, the model now returns the average speed and the average squared speed for each bin.  

The count rate calculation in \citet{sokol_etal:15b} was carried our using the scaling function that is constant for each ISN season:
\begin{equation}
S_\text{Y}(i, v, \{a_{2009},a_{2010},\dots,a_{2014}\}) = a_{y(i)},
\label{eq:defConstSFun}
\end{equation}
where $y(i)$ return the years that represent each of the ISN seasons. This form represents Y-norms and means that the scaling function does not depend on the atom impact speed. Inclusion of more ESA steps would require three separate constants for each ESA step and each ISN season. However, we verified that this is not necessary because the norms for individual seasons are constant within two periods: before and after the decrease of the PAC voltage. Consequently, we define the P-norms using the scaling function in the form:
\begin{equation}
S_\text{P}(i, v, \{a_{p,\text{ESA}e}\}_{p=\text{H,L};\,e=1,2,3}) = a_{p(i),\text{ESA}e(i)},
\label{eq:defSP}
\end{equation}
where $p(i)$ returns H or L for high/low PAC voltage, depending if a given data point $i$ was obtained before or after the decrease of the PAC voltage, and $e(i)$ returns the ESA step. Using the approach with P-norms implies that a different parameter $a$ is allowed for each energy step and the PAC voltage. As a result, for three energy steps the vector $\vec{q}$ of the parameters of function $S_\text{P}$ has 6 elements.

If one allows for a true dependence of the sensitivity of the instrument within a given ESA step on the atom speed, then the simplest possible form of the scaling function is the linear function $S_\text{PV}$, defined as:
\begin{align}
S_\text{PV}&(i, v, \{a_{p,\text{ESA}e},b_{p,\text{ESA}e}\}_{p=\text{H,L};\,e=1,2,3}) =\nonumber\\&= a_{p(i),\text{ESA}e(i)}(1 + b_{p(i),\text{ESA}e(i)}(v-v_0)),
\label{eq:defVSFun}
\end{align}
where $v_0$ is a certain reference speed, adopted here at 78~km~s$^{-1}$. This arbitrary choice does not affect the generality of the scaling function because it only requires a simple transformation for the parameters $a_{p,\text{ESA}e}$ for a different choice of $v_0$ . The function $S_{\text{PV}}$ may be regarded as an expansion of an unknown true sensitivity function into a Taylor series, cut off at the linear term. The parameters $b_{p,\text{ESA}e}$ are the relative change of the absolute sensitivity per km~s$^{-1}$. In general one may introduce a function $S_\text{YV}$ defined in analogy to $S_\text{PV}$ but with the parameters depending on the season, and not the H/L PAC voltage. 

\subsection{Parameter fitting}
\label{sec:fitting}

Parameter fitting is carried out using the method suggested by \citet{swaczyna_etal:15a} and successfully applied by \citet{bzowski_etal:15a} to obtain the inflow parameters of ISN He and  by \citet{kubiak_etal:16a} to obtain the inflow parameters of the Warm Breeze. \citet{swaczyna_etal:15a} discuss at length all data correlations that must be taken into account and we will not repeat all those details. We only briefly recall that the model must be calculated on a regular parameter grid and that one must compute the uncertainty covariance matrix $\tensor{V}$. With those on hand, we calculated the measure of goodness of fit given by $\chi^2$ defined as:
\begin{equation}
\chi^2\left( \vec{\pi}, \vec{q}\right) = \sum\limits_{i,j}^{} \left(c_i - g_{i}(\vec{\pi}, \vec{q})\right) \left(\tensor{V}^{-1}\right)_{i,j} \left(c_j - g_{j}(\vec{\pi}, \vec{q}) \right)
\label{eq:defChi2}
\end{equation}
where $i, j$ go over all data points used in the analysis, for all three ESA steps, $c_i$ is corrected count rate with the background and Warm Breeze contribution subtracted \citep{swaczyna_etal:15a}, and $g_i$ is calculated from Equation~\eqref{eq:defCntRate} with the $S$ functions given by Equations (\ref{eq:defConstSFun}--\ref{eq:defVSFun}). The fit parameters are $\vec{\pi}$ and $\vec{q}$. The parameters $\vec{\pi}$ correspond to the sought ISN parameters (the magnitude and ecliptic coordinates of the inflow velocity, and the temperature), and the parameters $\vec{q}$ are the unknown parameters of the sensitivity function $S$. The calculation was performed on sets of the ISN He parameters taken from a grid spaced at $\Delta \lambda = 1\degr$, $\Delta \beta = 0.1\degr$, $\Delta v = 0.2\text{ km s}^{-1}$, $\Delta v_{\mathrm{t}} = 0.1\text{ km s}^{-1}$, with 4607 nodes around the ISN He parameter values obtained by \citet{bzowski_etal:15a}. 

The optimum parameter set was found in a two-step process. The first step was finding the parameter set $(\vec{\pi}, \vec{q})$ with the minimum $\chi^2$ value calculated from Equation~\eqref{eq:defChi2}, evaluated for those 4607 parameter sets. Subsequently, a quadratic form in the parameter space was fitted to the $\chi^2$ surface near the intermediate minimum, and the parameter values for the absolute minimum of this surface were calculated analytically. This resulted in a better parameter resolution than that obtained directly from the parameter grid spacing. The fitting procedure is repeated for each considered data set and scaling function form. Consequently, both the inflow parameters and the parameters $\vec{q}$ vary accordingly.

Statistically, the expected value of $\chi^2_{\mathrm{min}}$ is given by $\chi^2_{\mathrm{min}} \simeq N_{\mathrm{dof}} \pm \sqrt{2 N_{\mathrm{dof}}}$, where $N_{\mathrm{dof}}$ is the number of degrees of freedom. For $N_{\mathrm{dof}}$ we adopted the number of data points minus the number of fit parameters, similarly to \citet{swaczyna_etal:15a, bzowski_etal:15a}, and \citet{kubiak_etal:16a}, even though this is strictly exact only for linear models, and our model is nonlinear. Finally, the uncertainties of the sought parameters are obtained from the curvature of the function $\chi^2(\vec{\pi},\vec{q})$ scaled using the scaling factor $S=\sqrt{\chi^2/N_\text{dof}}$ as presented in \citet{swaczyna_etal:15a} and \citet{bzowski_etal:15a}. The parameters $(\vec{\pi}, \vec{q})$ are correlated with each other as discussed in \citet{swaczyna_etal:15a}.

\section{Results}
\label{sec:results}

\begin{deluxetable*}{llccccccccc}
	\tablecaption{\label{tab:ISNparams}ISN He flow parameters}
	\tablewidth{0pt}
	\setlength{\tabcolsep}{2.5pt}
	\tablehead{
		\colhead{} & \colhead{Norm} & \colhead{ESA} &  \colhead{Seasons} & \colhead{$\lambda\ (\degr)$} & \colhead{$\beta\ (\degr)$} & \colhead{$v$ (km s$^{-1})$} & \colhead{$T$ (K)} & \colhead{$M$} & \colhead{$N_{\mathrm{dof}}$} & \colhead{$\frac{\chi^2}{N_{\mathrm{dof}}}$}
	}
	\startdata
0$^a$ & Y-norm& 2   & 2009--2014               & $255.74 \pm 0.45$ & $5.16 \pm 0.10$  & $25.76 \pm 0.37$ & $7440 \pm 260$  & $5.079 \pm 0.028$ & 254 & 1.84 \\
	1 & Y-norm 	& 2 	& 2009--2014	    	& $255.70 \pm 0.47$ & $5.14 \pm 0.11$  & $25.71 \pm 0.43$ & $7696 \pm 294$  & $4.980 \pm 0.020$ & 248 & 1.60 \\
	2 & Y-norm	& 2		& 2009--2015      		 & $255.68 \pm 0.46$ & $5.09 \pm 0.11$  & $25.65 \pm 0.39$ & $7677 \pm 285$  & $4.976 \pm 0.019$ & 289 & 1.55 \\
	3 & P-norm	& 2		& 2009--2015   			 		   & $255.46 \pm 0.45$ & $5.11 \pm 0.11$  & $25.85 \pm 0.40$ & $7805 \pm 278$  & $4.972 \pm 0.019$ & 294 & 1.61 \\
	4 & PV-norm	& 2		& 2009--2015             & $255.48 \pm 0.46$ & $4.91 \pm 0.09$  & $26.23 \pm 0.42$ & $7726 \pm 269$  & $5.072 \pm 0.025$ & 292 & 1.42 \\
	5 & P-norm	& 2--3	& 2009--2015             & $254.76 \pm 0.33$ & $5.20 \pm 0.08$  & $26.49 \pm 0.30$ & $8111 \pm 211$  & $4.999 \pm 0.015$ & 544 & 1.65 \\
	6 & PV-norm & 2--3	& 2009--2015             & $254.90 \pm 0.36$ & $5.04 \pm 0.08$  & $26.66 \pm 0.33$ & $7938 \pm 208$  & $5.085 \pm 0.019$ & 540 & 1.45 \\
	7 & P-norm	& 1--3	& 2009--2015             & $254.63 \pm 0.35$ & $5.23 \pm 0.07$  & $25.98 \pm 0.29$ & $7853 \pm 208$  & $4.982 \pm 0.016$ & 794 & 1.97 \\
	8 & PV-norm & 1--3	& 2009--2015             & $253.82 \pm 0.37$ & $5.06 \pm 0.07$  & $26.98 \pm 0.33$ & $8155 \pm 215$  & $5.077 \pm 0.020$ & 788 & 1.74 \\
%	\hline
	9 & P-norm	& 1--3	& 2009--$2015^{\dagger}$ & $255.62 \pm 0.36$ & $5.16 \pm 0.08$  & $25.82 \pm 0.33$ & $7673 \pm 225$  & $5.010 \pm 0.015$ & 686 & 1.64 \\ 
	10 & PV-norm	& 1--3	& 2009--$2015^{\dagger}$ & $255.41 \pm 0.40$ & $5.03 \pm 0.07$  & $26.21 \pm 0.37$ & $7691 \pm 230$  & $5.080 \pm 0.019$ & 680 & 1.49 \\
	\enddata
	\tablenotetext{a}{ results from \citet{bzowski_etal:15a}}
	\tablenotetext{\dagger}{data range restricted to longitudes within $115\degr$--$155\degr$ (see text)}
	
\end{deluxetable*}

\begin{deluxetable*}{lccccccc}
	\tablecaption{\label{tab:Ynorms}Y-norms fitted along with the ISN inflow parameters ($a_{y,\text{ESA}2}$, $10^{-6}~\text{cm}^2\text{sr}$)}
	\tablewidth{0pt}
	\tablehead{
		\colhead{} & \colhead{2009} & \colhead{2010} & \colhead{2011} & \colhead{2012} & \colhead{2013} & \colhead{2014} & \colhead{2015}}
	\startdata
0$^a$ & 16.76          &  16.66           &  15.11           &  14.50           &  7.709          &  7.977          &  $\cdots$ \\
	1 & $17.18 \pm 0.19$ & $17.08 \pm 0.20$ & $17.47 \pm 0.28$ & $17.27 \pm 0.22$ & $8.03 \pm 0.10$ & $8.25 \pm 0.10$ & $\cdots$ \\
	2 & $17.16 \pm 0.19$ & $17.07 \pm 0.20$ & $17.45 \pm 0.28$ & $17.25 \pm 0.22$ & $8.02 \pm 0.09$ & $8.25 \pm 0.10$ & $8.28 \pm 0.09$ \\
	\enddata
	\tablenotetext{a}{ results from \citet{bzowski_etal:15a}}
	\tablecomments{The row numbers correspond to the cases presented in Table~\ref{tab:ISNparams}}
\end{deluxetable*}

\begin{deluxetable*}{rcccccc}
	\tablecaption{\label{tab:Pnorms}P-norms and PV-norms fitted along with the ISN inflow parameters}
	\tablewidth{0pt}
	\tablehead{
		\colhead{}  & \colhead{H, ESA1} & \colhead{L, ESA1} & \colhead{H, ESA 2} & \colhead{L, ESA 2} & \colhead{H, ESA 3} & \colhead{L, ESA 3}
	}
	\startdata
	\hline
	\hline
		\multicolumn{7}{c}{Parametrs $a_{p,\text{ESA}e}$ ($10^{-6}~\text{cm}^2\text{sr}$)}\\
	\hline
	3 &   $\cdots$     &  $\cdots$     & $17.27\pm0.17$ & $8.12\pm0.09$ &    $\cdots$    &      $\cdots$     \\
	4 &   $\cdots$     &  $\cdots$     & $19.26\pm0.38$ & $8.63\pm0.14$ &    $\cdots$    &      $\cdots$     \\
	5 & $\cdots$ 	   & $\cdots$	   & $17.44\pm0.14$ & $8.25\pm0.06$ & $20.78\pm0.19$ & $10.22\pm0.08$  \\
	6 & $\cdots$ 	   & $\cdots$	   & $19.63\pm0.36$ & $8.80\pm0.16$ & $22.06\pm0.50$ & $10.35\pm0.20$  \\
	7 & $14.14\pm0.14$ & $6.29\pm0.05$ & $17.20\pm0.14$ & $8.18\pm0.06$ & $20.56\pm0.20$ & $10.12\pm0.08$  \\
	8 & $14.66\pm0.41$ & $7.09\pm0.16$ & $20.60\pm0.41$ & $9.31\pm0.18$ & $22.85\pm0.58$ & $10.96\pm0.23$  \\
	9 & $14.08\pm0.14$ & $6.27\pm0.06$ & $17.22\pm0.14$ & $8.13\pm0.07$ & $20.50\pm0.19$ & $10.09\pm0.08$  \\
	10 & $14.49\pm0.34$ & $6.57\pm0.13$ & $19.20\pm0.35$ & $8.63\pm0.15$ & $21.32\pm0.49$ & $10.20\pm0.19$  \\
	\hline
	\hline
		\multicolumn{7}{c}{Parametrs $b_{p,\text{ESA}e}$ ($10^{-2}~\text{km}^{-1}\text{s}$)}\\
	\hline
	4 &  $\cdots$    & $\cdots$     & $-5.8\pm0.8$ & $-3.7\pm0.8$ &  $\cdots$     &  $\cdots$    \\
	6 &  $\cdots$    & $\cdots$     & $-5.6\pm0.7$ & $-4.0\pm0.8$ &  $-3.3\pm0.9$ & $-1.6\pm0.9$ \\
	8 & $-1.9\pm1.0$ & $-5.5\pm0.9$ & $-4.2\pm0.9$ & $-6.0\pm0.8$ &  $-4.2\pm0.9$ & $-3.9\pm0.9$ \\
	10 & $-1.8\pm1.0$ & $-3.1\pm1.0$ & $-5.6\pm0.8$ & $-3.8\pm0.9$ &  $-2.3\pm1.0$ & $-1.2\pm1.0$ \\	
	\enddata
	\tablecomments{The row numbers correspond to the cases presented in Table~\ref{tab:ISNparams}.}
\end{deluxetable*}

We began the parameter evaluation by repeating the analysis of \citet{bzowski_etal:15a}. We used a slightly more restrictive good times list and slightly modified levels of the ubiquitous background \citep{galli_etal:16a, galli_etal:17a}. The contribution from the Warm Breeze was estimated from WTPM using the Warm Breeze parameters from \citet{kubiak_etal:16a}, i.e., different from the Warm Breeze estimate used by \citet{bzowski_etal:15a}. This contribution was subtracted from the data and the related uncertainty added to the uncertainty system. We performed fits assuming Y-norms and using data from observation seasons 2009--2014 and ESA step 2 only. The ISN He fit parameters are presented in row 1 in Table~\ref{tab:ISNparams}, and the corresponding Y-norms in Table \ref{tab:Ynorms}. In rows 0 of these tables the parameters found by \citet{bzowski_etal:15a} are presented for comparison. We conclude that the new data selection, as well as the new Warm Breeze subtraction does not substantially affect the ISN He inflow parameters. The differences are small, well within the uncertainties. The largest discrepancy is in the temperature ($\sim$250 K), and this is likely caused by the new Warm Breeze model subtracted here. The temperature of the Warm Breeze was found significantly lower by \citet{kubiak_etal:16a} than it was obtained from the earlier analysis \citep{kubiak_etal:14a}. \citet{bzowski_etal:15a} adopted the higher temperature of the Warm Breeze from \citet{kubiak_etal:14a}. Consequenty, after subtraction of the Warm Breeze in their analysis, the data featured a narrower peak, which the fitting procedure interpreted as due to a lower temperature of ISN He. At the same time we note that the reduced $\chi^2$ value, equal to $\chi^2/N_\text{dof}$, is significantly lower, i.e., the model better fits the data than in \citet{bzowski_etal:15a}.

In the next step, we added the data from season 2015 to the analysis (row 2). The resulting parameters remain almost the same. The inspection of Table~\ref{tab:Ynorms} shows that the obtained parameters of the scaling function are constant for the seasons with the H PAC voltage and separately with the L PAC voltage. The variation of these parameters is reduced by an order of magnitude with respect to the one obtained by \citet{bzowski_etal:15a}. Moreover, the new values agree in these two periods, and the usage of the P-norms rather than Y-norms is justified. Results for the P-norms applied to ESA step 2 solely are presented in row 3 of Tables~\ref{tab:ISNparams} and~\ref{tab:Pnorms}. The obtained values of the norms agree with the values formerly found for each season. The ISN flow parameters are slightly different, but well within the uncertainty ranges. 

We found that including data from the 2016 campaign significantly increases the value of the reduced $\chi^2$. During this season, only data from three orbits before the absolute maximum of ISN He count rate could be obtained due to a temporary issue with the IBEX star tracker system, which makes the data coverage for this campaign significantly poorer. Therefore, we decided to not include the data from this season in this analysis.

Next, we checked how the fit parameter values change when a variation of the instrument sensitivity with atom speed is allowed for, while continuing to use data solely from ESA step 2. We used the function $S_\text{PV}$ from Equation~\eqref{eq:defVSFun} and list the results in row 4 of Table~\ref{tab:ISNparams}. This modification resulted in a significant reduction of the reduced $\chi^2$ value. The inflow latitude and speed changed appreciably but within the uncertainty range. 

Subsequently, we tested whether including additional ESA steps while assuming no dependence of the instrument sensitivity to atom speed within individual ESA step returns similar results to those obtained by \citet{bzowski_etal:15a}, using again the sensitivity function $S_{\mathrm{P}}$. Finally, we allowed for the sensitivity within an ESA step to vary with atom speed while continuing to use data from three ESA steps. To that end, we took PV-norms and adopted the sensitivity function $S_\text{PV}$ from Equation~\eqref{eq:defVSFun}. The resulting ISN He parameters are presented in rows 5--8 of Table~\ref{tab:ISNparams}, and the norms in Table~\ref{tab:Pnorms}. As a result of these fittings we found that the ISN flow parameters are substantially different if all three ESA steps are included (rows 7 and 8) from those obtained solely for ESA step 2, especially for the inflow longitude. This effect is not present if ESA step 1 is excluded (rows 5 and 6). The simultaneous increase in the reduced $\chi^2$ value for the three ESA steps suggests that the model fit is significantly poorer. This is likely due to a contribution from ISN H, unaccounted for in the model. We consider this hypothesis is likely because eliminating ESA step 1 from the data and fitting to data from ESA steps 2 and 3 results in parameters similar to those obtained previously (i.e., no important change compared to \citet{bzowski_etal:15a}), and our analysis shows that the presence of ISN H is the most prominent in ESA step 1, consistently with the findings by \citet{saul_etal:12a} and \citet{schwadron_etal:13a}. 

Most of the ISN H signal can be avoided if one adopts a tighter restriction on the spin axis orientation, which corresponds to the ecliptic longitude of IBEX. Here, we limited them to longitudes in the range $(115\degr, 155\degr)$ compared to $(115\degr, 160\degr)$ used so far. Effectively, this removed data from those orbits where the expected contribution from ISN H was the largest within the dataset. As a result, we noticed that this restriction improved the fit quality as it reduced the reduced $\chi^2$ value by $\sim$0.3 both for the case with P-norms and for that with PV-norms. This restriction effectively resulted in leaving out 6 late orbits, i.e., those well after the peak of ISN He is observed (orbits 19, 67, 115, 157b, 158a, 238a). The contribution from ISN H to the signal observed during these orbits is expected to be the largest during a given ISN season. Therefore, as the final result for the ISN He parameters we adopt those listed in rows 9 and 10 in Table~\ref{tab:ISNparams}, with the parameters of the scaling functions in Table~\ref{tab:Pnorms}. 

The obtained parameters of the scaling functions suggest that the efficiency of IBEX-Lo is increasing within the three ESA steps. Moreover, the high-to-low PAC voltage ratios of the  parameters $a_{\text{H,ESA}e}/a_{\text{L,ESA}e}$ are similar for all three ESA steps (2.21, 2.22, and 2.09 for ESA steps 1, 2, and 3, respectively, from the case presented in row 10 in Table~\ref{tab:Pnorms}). 

\section{Discussion and conclusions}
\label{sec:discussion}

The differences between the present analysis of the bulk velocity and temperature of ISN He and that carried out by \citet{bzowski_etal:15a} are the following: (1) subtracting from the data of a better model of the Warm Breeze (from \citet{kubiak_etal:16a} instead of that from \citet{kubiak_etal:14a}); (2) adopting more stringent criteria for good times and using slightly different background level for the observation seasons after 2012 \citep{galli_etal:16a}; (3) using P-norms instead of Y-norms (i.e., adopting instrument sensitivity parameters characteristic for a given PAC voltage rather than for a given observation season); (4) using data from one more ISN observation season (2015) while cutting off some late orbits during each season; (5) using data from ESA steps 1 through 3 instead of only ESA step 2; (6) allowing  for the instrument sensitivity to depend on the impact speed of He atoms within individual ESA steps (function $S_\text{PV}$ from Equation~\ref{eq:defVSFun} instead of a constant value).

\begin{figure} 
%\epsscale{1.15} 
\epsscale{.7} 
\plotone{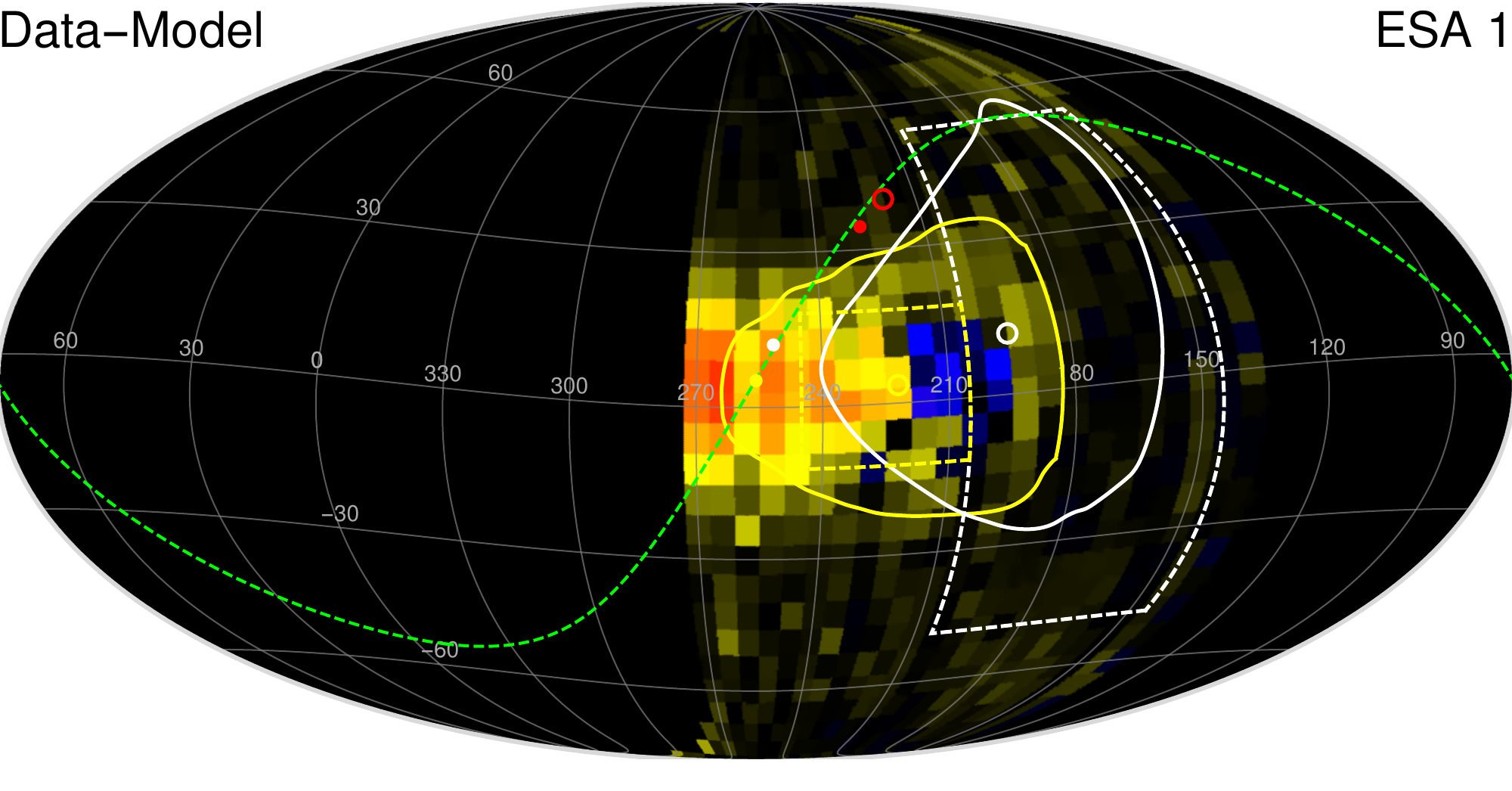} 
\plotone{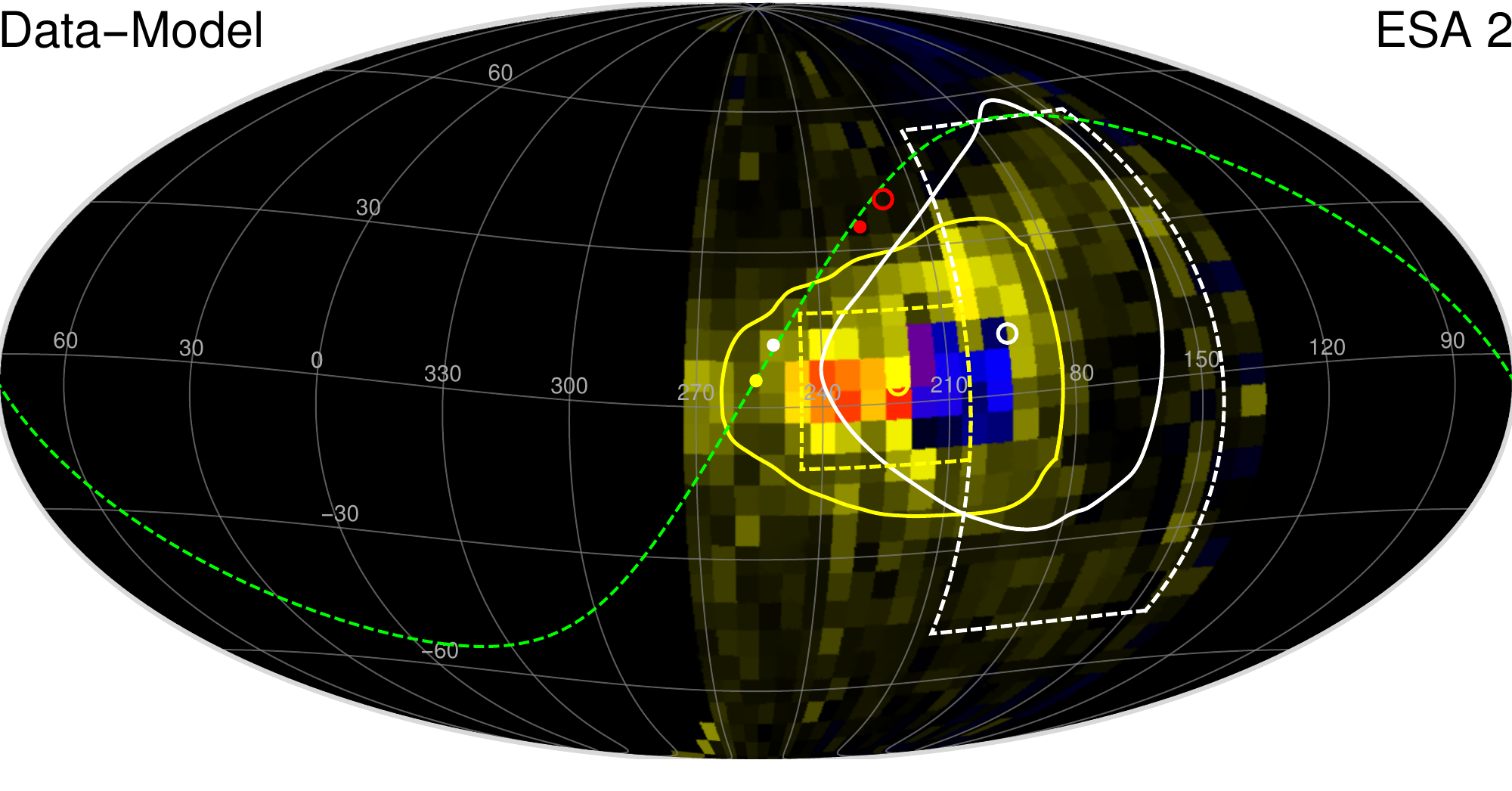} 
\plotone{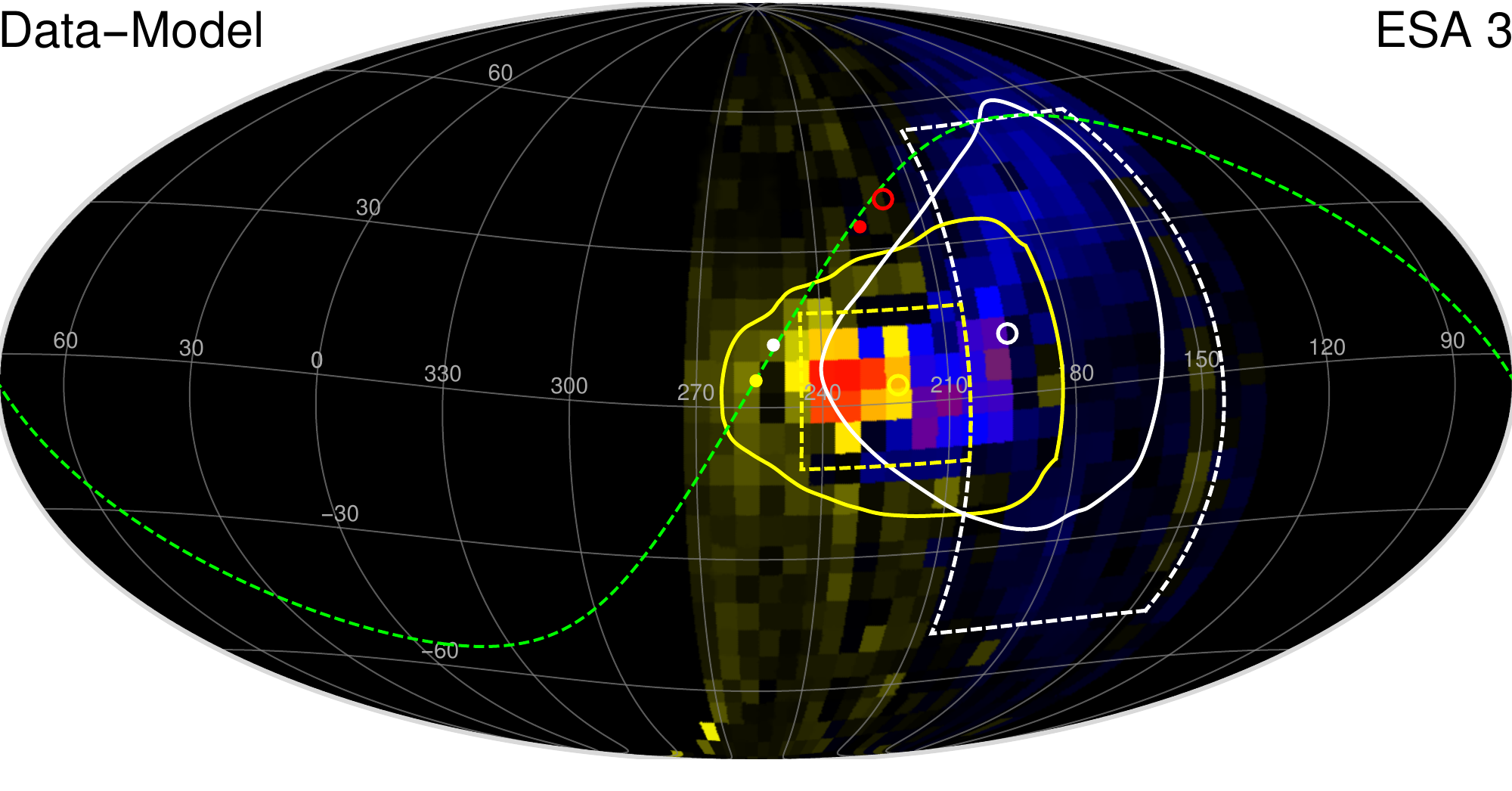}
\plotone{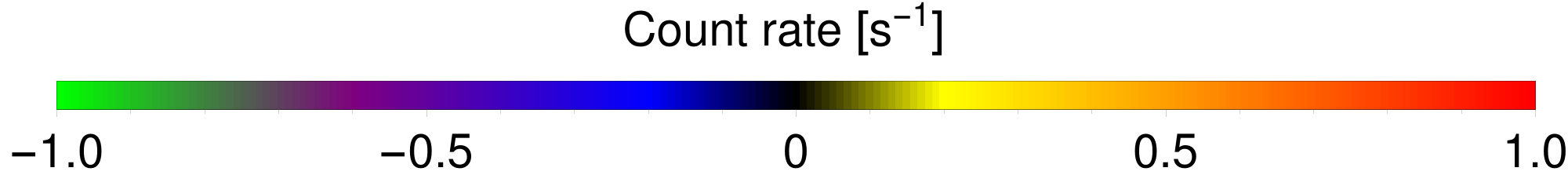} 
\caption{Sky maps of the residuals (data minus model), weighted-averaged over all ISN observation seasons (2009--2015), shown as count rate in s$^{-1}$ in ESA steps 1 (top), 2 (middle), and 3 (bottom) in the J2000 ecliptic coordinates in the IBEX-inertial frame. The lines, points, and contours are identical to those in Figure~\ref{fig:map`E123}. 
}
\label{fig:map`resid}
\end{figure}

\begin{figure}
%\epsscale{1.15} 
\epsscale{.7} 
\plotone{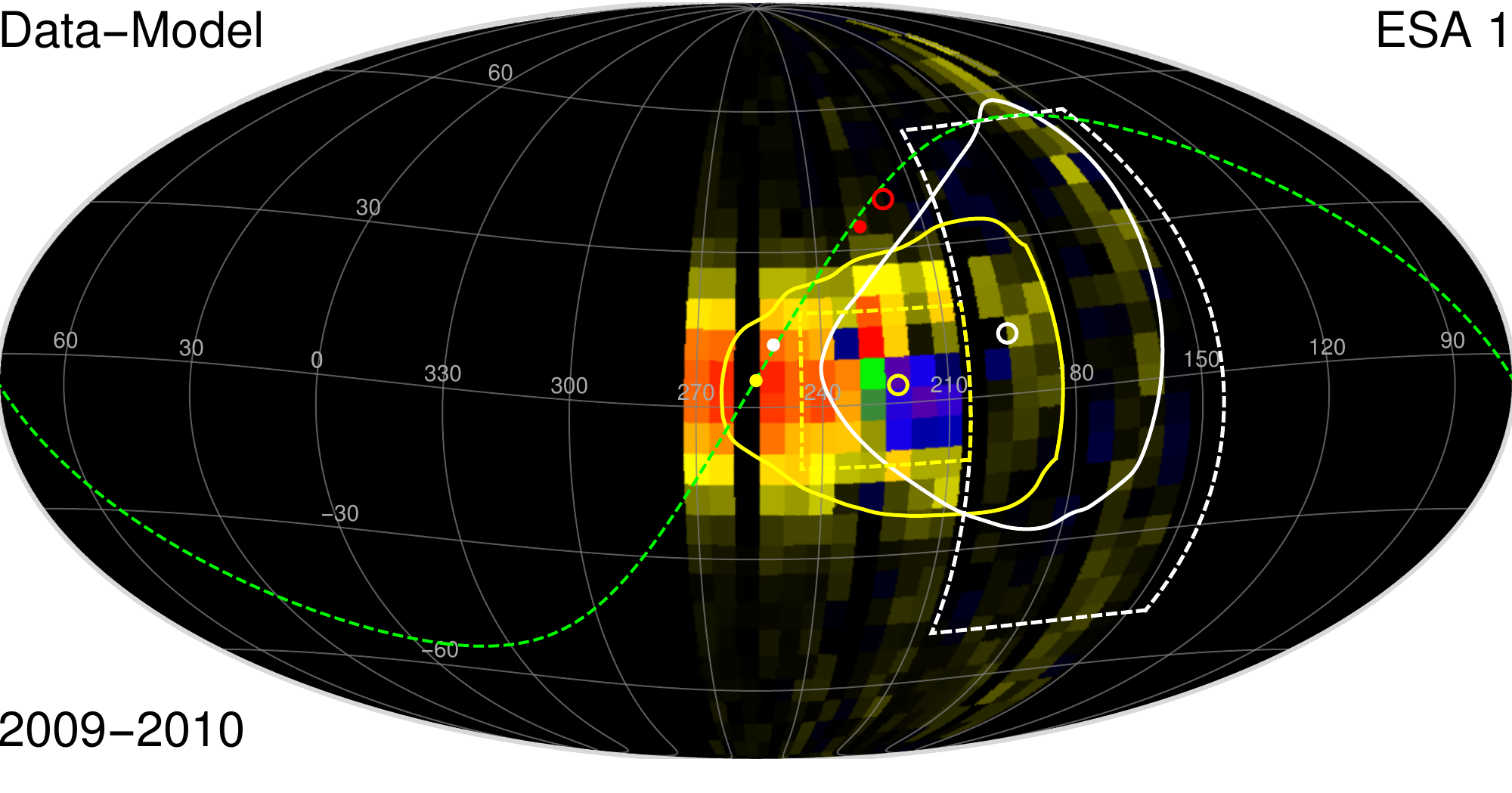} 
\plotone{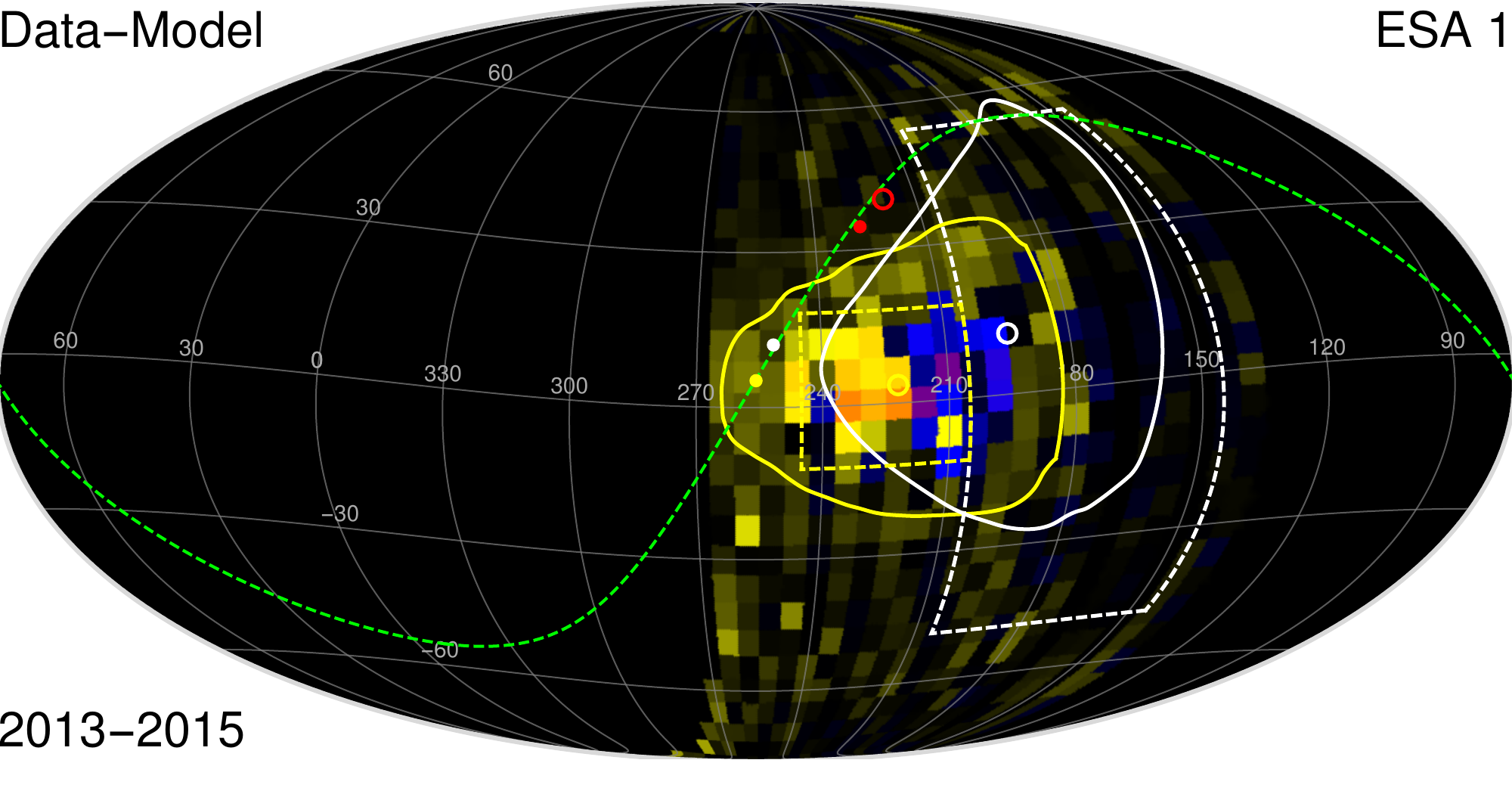} 
\plotone{legend_residuals.pdf} 
\caption{Sky maps of the residuals (data $-$ model) for ESA step 1, averaged over ISN observation seasons 2009 and 2010 from the minimum of solar activity (upper panel) and over ISN seasons 2013 through 2015 from the maximum of solar activity (lower panel). The lines, points, and contours are identical to those in Figure~\ref{fig:map`E123}}
\label{fig:map`e1`residMinMax}
\end{figure}

Modifications (1) and (2) result in a small change in the Mach number of the flow, reducing it from $5.079 \pm 0.028$ obtained by \citet{bzowski_etal:15a} to $4.980 \pm 0.020$. This reduction is mostly due to an increase in the fitted temperature by 230~K. While not important for the global image of the heliosphere, this is a statistically significant difference in the Mach number of more than 3.5$\sigma$. Modifications of the direction and speed of inflow and of the temperature of the ISN He are within 1--2$\sigma$. In this respect, switching from Y-norms to P-norms and using data from one more observation seasons (modifications 3 and 4), as well as using data from three ESA steps instead of one (5), do not result in any further statistically significant modifications of the ISN He parameters. Adding more data and applying more stringent filtering results in a certain reduction of $\chi^2$ per degree of freedom, but still these $\chi^2$ significantly exceed the expected value. The small increase in the temperature is understandable since the temperature of the Warm Breeze obtained by \citet{kubiak_etal:16a} was lower than that obtained by \citet{kubiak_etal:14a}. As a result, the change in the modeled Warm Breeze signal subtracted from the data was the largest for the bins at spin angles farthest from the peaks, which resulted in a larger signal in these bins left in the ISN He signal used in parameter fitting.

Allowing for the instrument sensitivity within an ESA step to vary with the impact speed (modification 6) restores the Mach number to the value originally obtained by \citet{bzowski_etal:15a}. Also, the other parameters are little affected in comparison with those obtained by \citet{bzowski_etal:15a}, even though the latitude of the flow is slightly lower, and the temperature and the speed are slightly larger. From this analysis we conclude that the current best estimate for the bulk velocity vector and temperature of ISN He are those listed in the last two rows of Table~\ref{tab:ISNparams}. These two parameter sets are in good agreement. The difference between them is presented in  Table~\ref{tab:ISNparamsComp}. This table shows also a comparison with the results of the previous studies of the IBEX data. The parameters obtained in this analysis are in agreement with the "working values" provided by \citet{mccomas_etal:15b} based on the analyses by \citet{bzowski_etal:15a} and \citet{schwadron_etal:15a}. 

\begin{deluxetable*}{lcccc}
	\tablecaption{\label{tab:ISNparamsComp}ISN He flow parameters from different studies of IBEX data}
	\tablewidth{0pt}
	\setlength{\tabcolsep}{2.5pt}
	\tablehead{
		\colhead{Reference} & \colhead{$\lambda\ (\degr)$} & \colhead{$\beta\ (\degr)$} & \colhead{$v$ (km s$^{-1})$} & \colhead{$T$ (K)} }
	\startdata
	This analysis P-norm (9)	& $255.62 \pm 0.36$ & $5.16 \pm 0.08$  & $25.82 \pm 0.33$ & $7673 \pm 225$  \\ 
	This analysis PV-norm (10)  & $255.41 \pm 0.40$ & $5.03 \pm 0.07$  & $26.21 \pm 0.37$ & $7691 \pm 230$  \\
	Difference (9) $-$ (10) 	    & $0.21$			& $0.13$		   & $-0.39$	  	  & $-18$			\\
	\hline
	\citet{bzowski_etal:15a} &  $255.74 \pm 0.45$ & $5.16 \pm 0.10$  & $25.76 \pm 0.37$ & $7440 \pm 260$  \\
	\citet{schwadron_etal:15a} & $255.6\pm1.4$ & $5.12\pm0.27$ & $25.4\pm1.1$ & $8000\pm1300$ \\
	\citet{mccomas_etal:15b}\tablenotemark{a}  & 255.7 & 5.1 & 25.4 & 7500\\
	\enddata
	\tablenotetext{a}{ "Working values" based on \citet{bzowski_etal:15a} and \citet{schwadron_etal:15a}}
\end{deluxetable*}

The combination of the model of ISN He obtained now and of the Warm Breeze model from \citet{kubiak_etal:16a} is still not perfect. The most significant issue is likely the presence of ISN H in the data. This hypothesis is based on analysis of residuals, i.e., difference between the count rates in the data and the model. Maps of these residuals, averaged over all observation seasons, are presented in Figure~\ref{fig:map`resid} separately for the three ESA steps. The yellow, orange, and red colors mark positive differences, i.e., an excess of the data over the model. As clearly seen, this excess is mostly located in the region where ISN H is expected based on modeling of the ISN H signal \citep[see, e.g., ][]{kubiak_etal:13a}. This is also in qualitative agreement with data analyses in \citet{saul_etal:12a, saul_etal:13a} and \citet{schwadron_etal:13a}. Both the magnitude and spatial range of the excess are reduced when going from ESA steps 1 to 3 and they evolve with time, as is clearly implied from a comparison of the map of the residuals for ESA step 1, ISN observation seasons 2009 and 2010, shown in the upper panel of Figure~\ref{fig:map`e1`residMinMax}, with those obtained for ISN seasons 2013 through 2015, shown in the lower panel of this figure. 

It is important to note that the residual excess is present at all times, which suggests that it is indeed due to the physical reason just discussed and not necessarily to an inadequacy of the adopted model of the ISN He and Warm Breeze. Furthermore, the intensity of the excess qualitatively conforms with the expectations for the evolution of ISN H flux during the solar cycle, with a stronger signal expected during low solar activity and a reduction in the signal strength when the activity increases \citep{kubiak_etal:13a}. In our case, the excess for the ISN seasons 2009 and 2010 is larger than that for the solar maximum seasons 2013--2015. This is understandable because during 2009 and 2010 the Sun had just emerged from a prolonged minimum of activity when both solar radiation pressure acting on interstellar H and the ionization losses were the lowest. By contrast, during 2013--2015 ISN seasons the Sun was at the maximum of its activity and both the high ionization rate \citep[see Figure~31 in][]{mccomas_etal:17a} and high radiation pressure reduced the flux of ISN H at 1~au considerably. As a result, the flux of ISN H at 1~au during this time is expected to be lower than during the minimum of solar activity. 

The magnitude of the residuals likely underestimates the true magnitude of the ISN H contribution to the observed signal. This is because the fitting procedure aims to fit the data to the model as good as possible within the adopted model, which in our case does not include ISN H. The fitting procedure we have used does not take into account possible systematic patterns in the residuals. In the case of a perfect fit, with all signal components taken into account, the spatial distribution of positive and negative residual values should be random. In our case, evidently connected regions of negative and positive residuals exist. The region of positive residuals extends well beyond the data subset used in the fitting. These are indications that the model, which only includes the primary ISN He population and the Warm Breeze, is missing an extra component, which is likely ISN H. Since, however, the optimization procedure seeks to have the  mean value of the residuals close to 0, some counts due to ISN H were inadvertently attributed to ISN He. Therefore it is not recommended to adopt the residuals as a signal from ISN H at face value. The true contributions from ISN H must be assessed separately, based on a model that contains all components ISN H, ISN He, and the Warm Breeze. Moreover, the residuals to the right of the ISN He peak show a cluster of pixels with negative values surrounded by pixels with positive values in ESA steps 1 and 2. The same region shows predominantly negative residuals in ESA step 3. It suggests that also the model of the Warm Breeze needs to be revised.

The presence of neutral H in the signal during all ISN observation seasons suggests that before further analysis of details of ISN He and the Warm Breeze the question of ISN H must be addressed. This topic was discussed by \citet{schwadron_etal:13a} and \citet{katushkina_etal:15b}, who pointed out that the ratio of counts from ISN H actually observed in ESA steps 1 and 2 is significantly different from that expected from state of the art models of ISN H in the heliosphere and that the likely culprit is an inadequate knowledge of the solar Lyman-$\alpha$ radiation pressure, in particular of the spectral profile of this line. Analysis of the largest currently available data set of full-disk solar Lyman-$\alpha$ line profiles by \citet{kowalska-leszczynska_etal:17a} (43 profiles measured during various phases of solar activity) showed that indeed, the central reversal of the profile is deeper than that used by \citet{katushkina_etal:15b}, who adopted a model by \citet{tarnopolski_bzowski:09} which had been developed based on just 9 profiles then available. Analysis of the implications of this finding for the ISN H signal expected from IBEX-Lo is underway now. 

The question of the sensitivity of IBEX-Lo to He atoms in various ESA steps is interesting mostly in the studies of the secondary population of ISN He, i.e., the Warm Breeze. Studying the energy spectrum of the Warm Breeze may potentially bring a better insight into the plasma flow and temperature in the outer heliosheath. Also, as shown by our results, expanding the analysis of ISN He from one to three ESA steps improves the statistics. Additionally, one obtains a better opportunity to study ISN H, which is the most abundant in ESA step 1. Therefore understanding the differences in the sensitivity of IBEX-Lo to neutral He in the lowest-energy ESA steps is important. 

Our analysis supports the conclusion by \citet{bzowski_etal:15a} and \citet{mobius_etal:15b} that reducing the PAC voltage after the 2012 ISN observation season resulted in a reduction of the overall sensitivity of IBEX-Lo to He atoms  by  factor of $\sim$2.2 for ESA steps 1 and 2, and by a factor of $\sim$2.1 for ESA step 3. These reductions were taken into account when calculating the averaged maps shown in Figures~\ref{fig:map`E123}, \ref{fig:map`resid}, and \ref{fig:map`e1`residMinMax}.

An interesting and surprising finding is that the slope of the speed dependence of sensitivity within the energy steps is negative, i.e., the sensitivity is larger for {\em{slower}} speeds of the incoming neutral atoms. This speed dependence came out the largest for ESA step 2, consistently for both PAC voltage magnitudes used. The value of $b\sim5\cdot10^{-2}$~$\text{km}^{-1}\text{s}$, typical for ESA step 2 (see Table~\ref{tab:Pnorms}), corresponds to a change of the scaling function by $\sim$25\% for the range of the mean velocity of incoming atoms in the selected pixels of $\Delta v\sim5~\text{km}~\text{s}^{-1}$. This behavior is contrary to intuition, especially when one recalls that \citet{kubiak_etal:14a}, \citet{galli_etal:15a}, and \citet{sokol_etal:15a} detected a lower energy threshold for the sensitivity in ESA step 2. Whether this is indeed a behavior of the sputtered ion distribution or by how much this effect is produced by the presence of ISN H in the observations, still needs to be evaluated. We believe that either there must be a sensitivity maximum for a certain unknown atom speed within an ESA step, which is missed due to the adoption of the linear sensitivity functions $S_{\text{PV}}(i,v)$ or it is an artifact resulting from the presence of an unaccounted contribution from ISN H in the signal. 

It is important to realize that the speed sensitivity within the ESA steps was established only for the largest speed values of neutral He (75--81~km~s$^{-1}$), characteristic for the spin angle bins with the largest count rate in the orbits used in the analysis. The other pixels, as well as those occupied by the signal from the Warm Breeze, have lower characteristic speeds (down to $\sim$60~km~s$^{-1}$; see Figure~\ref{fig:map`speeds}). Therefore extrapolating the sensitivity functions resulting from our analysis significantly downward in speed is strongly discouraged. Free from this issue are the parameters obtained assuming that the sensitivity within ESA steps is independent of energies. 

Further investigation of the primary and secondary populations of ISN He must be carried out in parallel with analysis of the ISN H component in the data, in an iterative way. As a first step, the contribution from ISN H should be tentatively identified and subtracted from the data. With this, one of the parameter sets from the fits we have obtained in this paper should be adopted and the model signal due to ISN He should be computed and subtracted from the data. Subsequently, a new parameter set for the Warm Breeze fitted should be fitted to the resulting data set, and the resulting Warm Breeze model used in the next iteration of parameter fitting for ISN He and ISN H. 

\section{Summary}
\label{sec:summary}
We have analyzed observations of ISN He carried out by IBEX during the ISN He observation seasons 2009 through 2015, for the first time using information from IBEX-Lo ESA steps 1 through 3. We have established the differences in the sensitivity to ISN He atoms in these ESA steps and the sensitivity change due to the change in the IBEX-Lo PAC voltage setting that was introduced after the 2012 ISN season. We found that the overall sensitivity increases from lower energy steps to the higher ones. Surprisingly, however, we found that the sensitivity within all these ESA steps may be a {\em{decreasing}} function of atom speed. We are currently unable to verify if this is a true effect or a result of the presence of counts due to ISN H in the signal. 

The excesses of the measured count rates over the modeled signal on the left side of the ISN He peak position in all ESA steps suggests that the data may be partially contaminated by ISN H atoms. We suggest that the signal due to ISN H must be quantitatively interpreted and subtracted before information on the Warm Breeze, available in the measurements in ESA steps 1--3, can be fully utilized. Finally, we found that the bulk velocity vector and temperature of ISN He obtained in the present analysis using data from ESA steps 1--3, listed in the last two rows in Table~\ref{tab:ISNparams}, are within the fit uncertainties of these parameters obtained by \citet{bzowski_etal:15a} from the analysis using only data from ESA step 2. Further analysis of the primary and secondary populations of ISN He and ISN H should be carried out iteratively, using parameters of these populations obtained in a preceding step of the analysis in the subsequent steps.

\acknowledgments
This study was supported by the grant 2015/18/M/ST9/00036 from the National Science Center, Poland. P.S. is supported by the Foundation for Polish Science (FNP). Work in the USA and the IBEX data were supported by the IBEX mission as a part of NASA's Explorer Program, grant NNG05EC85C.

\bibliographystyle{aasjournal}
\bibliography{iplbib}{}

\begin{thebibliography}{}
\expandafter\ifx\csname natexlab\endcsname\relax\def\natexlab#1{#1}\fi
\providecommand{\url}[1]{\href{#1}{#1}}

\bibitem[{Baranov \& Malama(1995)}]{baranov_malama:95}
Baranov, V.~B., \& Malama, Y.~G. 1995, \aap, 304, 14755

\bibitem[{{Bzowski} {et~al.}(2017){Bzowski}, {Kubiak}, {Czechowski}, \&
  {Grygorczuk}}]{bzowski_etal:17a}
{Bzowski}, M., {Kubiak}, M.~A., {Czechowski}, A., \& {Grygorczuk}, J. 2017,
  \apj, 845, 15

\bibitem[{{Bzowski} {et~al.}(2014){Bzowski}, {Kubiak}, {H{\l}ond},
  {Sok{\'o}{\l}}, {Banaszkiewicz}, \& {Witte}}]{bzowski_etal:14a}
{Bzowski}, M., {Kubiak}, M.~A., {H{\l}ond}, M., {et~al.} 2014, \aap, 569, A8

\bibitem[{Bzowski {et~al.}(2012)Bzowski, Kubiak, M{\"o}bius, Bochsler, Leonard,
  Heirtzler, Kucharek, Sok{\'{o}}{\l}, H{\l}ond, Crew, Schwadron, Fuselier, \&
  McComas}]{bzowski_etal:12a}
Bzowski, M., Kubiak, M.~A., M{\"o}bius, E., {et~al.} 2012, \apjs, 198, 12

\bibitem[{Bzowski {et~al.}(2013)Bzowski, Sok{\'{o}}{\l}, Tokumaru, Fujiki,
  Qu{\'e}merais, Lallement, Ferron, Bochsler, \& McComas}]{bzowski_etal:13a}
Bzowski, M., Sok{\'{o}}{\l}, J.~M., Tokumaru, M., {et~al.} 2013, in
  {Cross-Calibration of Far {UV} Spectra of Solar Objects and the Heliosphere},
  ed. E.~Qu{\'e}merais, M.~Snow, \& R.~Bonnet, {ISSI Scientific Report} No.~13
  ({Springer Science+Business Media}), 67--138

\bibitem[{{Bzowski} {et~al.}(2015){Bzowski}, {Swaczyna}, {Kubiak},
  {Sok\'{o}{\l}}, {Fuselier}, {Galli}, {Heirtzler}, {Kucharek}, {Leonard},
  {McComas}, {M{\"o}bius}, {Schwadron}, \& {Wurz}}]{bzowski_etal:15a}
{Bzowski}, M., {Swaczyna}, P., {Kubiak}, M., {et~al.} 2015, \apjs, 220, 28

\bibitem[{{Frisch} {et~al.}(2011){Frisch}, {Redfield}, \&
  {Slavin}}]{frisch_etal:11a}
{Frisch}, P.~C., {Redfield}, S., \& {Slavin}, J.~D. 2011, \araa, 49, 237

\bibitem[{{Funsten} {et~al.}(2013){Funsten}, {DeMajistre}, {Frisch},
  {Heerikhuisen}, {Higdon}, {Janzen}, {Larsen}, {Livadiotis}, {McComas},
  {M{\"o}bius}, {Reese}, {Reisenfeld}, {Schwadron}, \&
  {Zirnstein}}]{funsten_etal:13a}
{Funsten}, H.~O., {DeMajistre}, R., {Frisch}, P.~C., {et~al.} 2013, \apj, 776,
  30

\bibitem[{{Fuselier} {et~al.}(2009){Fuselier}, {Bochsler}, {Chornay}, {Clark},
  {Crew}, {Dunn}, {Ellis}, {Friedmann}, {Funsten}, {Ghielmetti}, {Googins},
  {Granoff}, {Hamilton}, {Hanley}, {Heirtzler}, {Hertzberg}, {Isaac}, {King},
  {Knauss}, {Kucharek}, {Kudirka}, {Livi}, {Lobell}, {Longworth}, {Mashburn},
  {McComas}, {M{\"o}bius}, {Moore}, {Moore}, {Nemanich}, {Nolin}, {O'Neal},
  {Piazza}, {Peterson}, {Pope}, {Rosmarynowski}, {Saul}, {Scherrer}, {Scheer},
  {Schlemm}, {Schwadron}, {Tillier}, {Turco}, {Tyler}, {Vosbury}, {Wieser},
  {Wurz}, \& {Zaffke}}]{fuselier_etal:09b}
{Fuselier}, S.~A., {Bochsler}, P., {Chornay}, D., {et~al.} 2009, \ssr, 146, 117

\bibitem[{Fuselier {et~al.}(2012)Fuselier, Allegrini, Bzowski, Funsten,
  Ghielmetti, Gloeckler, Heirtzler, {P. Janzen}, Kubiak, Kucharek, McComas,
  Möbius, Moore, Petrinec, Quinn, {D. Reisenfeld}, Saul, Scheer, Schwadron,
  Trattner, Vanderspek, \& Wurz}]{fuselier_etal:12a}
Fuselier, S.~A., Allegrini, F., Bzowski, M., {et~al.} 2012, ApJ, 754, 14

\bibitem[{{Galli} {et~al.}(2015){Galli}, {Wurz}, {Park}, {Kucharek},
  {M{\"o}bius}, {Schwadron}, {Sok\'{o}{\l}}, {Bzowski}, {Kubiak}, {Swaczyna},
  {Fuselier}, \& {McComas}}]{galli_etal:15a}
{Galli}, A., {Wurz}, P., {Park}, J., {et~al.} 2015, \apjs, 220, 30

\bibitem[{Galli {et~al.}(2016)Galli, Wurz, Schwadron, Kucharek, M{\"o}bius,
  Bzowski, Sok{\'o}{\l}, Kubiak, Funsten, Fuselier, \&
  McComas}]{galli_etal:16a}
Galli, A., Wurz, P., Schwadron, N., {et~al.} 2016, \apj, 821, 107

\bibitem[{Galli {et~al.}(2017)Galli, Wurz, Schwadron, Kucharek, M{\"o}bius,
  Bzowski, Sok{\'o}{\l}, Kubiak, Fuselier, Funsten, \&
  McComas}]{galli_etal:17a}
---. 2017, \apj, 851, 2

\bibitem[{{Katushkina} {et~al.}(2015){Katushkina}, {Izmodenov}, \&
  {Alexashov}}]{katushkina_etal:15a}
{Katushkina}, O.~A., {Izmodenov}, V.~V., \& {Alexashov}, D.~B. 2015, \mnras,
  446, 2929

\bibitem[{Katushkina {et~al.}(2015)Katushkina, Izmodenov, Alexashov, Schwadron,
  \& McComas}]{katushkina_etal:15b}
Katushkina, O.~A., Izmodenov, V.~V., Alexashov, D.~B., Schwadron, N.~A., \&
  McComas, D.~J. 2015, \apjs, 220, 33

\bibitem[{{Kowalska-Leszczynska} {et~al.}(2017){Kowalska-Leszczynska},
  {Bzowski}, {Sok{\'o}{\l}}, \& {Kubiak}}]{kowalska-leszczynska_etal:17a}
{Kowalska-Leszczynska}, I., {Bzowski}, M., {Sok{\'o}{\l}}, J.~M., \& {Kubiak},
  M.~A. 2017, ApJ, submitted, arXiv:1710.06602

\bibitem[{Kubiak {et~al.}(2013)Kubiak, Bzowski, Sok{\'o\l}, M{\"o}bius,
  Rodr{\'i}guez, Wurz, \& McComas}]{kubiak_etal:13a}
Kubiak, M.~A., Bzowski, M., Sok{\'o\l}, J.~M., {et~al.} 2013, \aap, 556, A39

\bibitem[{{Kubiak} {et~al.}(2014){Kubiak}, {Bzowski}, {Sok{\'o}{\l}},
  {Swaczyna}, {Grzedzielski}, {Alexashov}, {Izmodenov}, {Moebius}, {Leonard},
  {Fuselier}, {Wurz}, \& {McComas}}]{kubiak_etal:14a}
{Kubiak}, M.~A., {Bzowski}, M., {Sok{\'o}{\l}}, J.~M., {et~al.} 2014, \apjs,
  213, 29

\bibitem[{Kubiak {et~al.}(2016)Kubiak, Swaczyna, Bzowski, Sok{\'o}{\l},
  Fuselier, Galli, Heirtzler, Kucharek, Leonard, McComas, Park, Schwadron, \&
  Wurz}]{kubiak_etal:16a}
Kubiak, M.~A., Swaczyna, P., Bzowski, M., {et~al.} 2016, \apjs, 223, 35

\bibitem[{Leonard {et~al.}(2015)Leonard, M{\"o}bius, Bzowski, Fuselier,
  Heirtzler, Kubiak, Kucharek, Lee, McComas, Schwadron, \&
  Wurz}]{leonard_etal:15a}
Leonard, T.~W., M{\"o}bius, E., Bzowski, M., {et~al.} 2015, \apj, 804, 42

\bibitem[{{McComas} {et~al.}(2015{\natexlab{a}}){McComas}, {Bzowski}, {Frisch},
  {Fuselier}, {Kubiak}, {Kucharek}, {Leonard}, {M{\"o}bius}, {Schwadron},
  {Sok{\'o}{\l}}, {Swaczyna}, \& {Witte}}]{mccomas_etal:15a}
{McComas}, D., {Bzowski}, M., {Frisch}, P., {et~al.} 2015{\natexlab{a}}, \apj,
  801, 28

\bibitem[{{McComas} {et~al.}(2015{\natexlab{b}}){McComas}, {Bzowski}, {Frisch},
  {Galli}, {Izmodenov}, {Katushkina}, {Kubiak}, {Lee}, {Leonard}, {M{\"o}bius},
  {Park}, {Schwadron}, {Sok{\'o}{\l}}, {Swaczyna}, {Wood}, \&
  {Wurz}}]{mccomas_etal:15b}
{McComas}, D., {Bzowski}, M.~{Fuselier}, S., {Frisch}, P., {et~al.}
  2015{\natexlab{b}}, \apjs, 220, 22

\bibitem[{{McComas} {et~al.}(2009){McComas}, {Allegrini}, {Bochsler},
  {Bzowski}, {Collier}, {Fahr}, {Fichtner}, {Frisch}, {Funsten}, {Fuselier},
  {Gloeckler}, {Gruntman}, {Izmodenov}, {Knappenberger}, {Lee}, {Livi},
  {Mitchell}, {M{\"o}bius}, {Moore}, {Pope}, {Reisenfeld}, {Roelof},
  {Scherrer}, {Schwadron}, {Tyler}, {Wieser}, {Witte}, {Wurz}, \&
  {Zank}}]{mccomas_etal:09a}
{McComas}, D.~J., {Allegrini}, F., {Bochsler}, P., {et~al.} 2009, \ssr, 146, 11

\bibitem[{{McComas} {et~al.}(2011){McComas}, {Carrico}, {Hautamaki},
  {Intelisano}, {Lebois}, {Loucks}, {Policastri}, {Reno}, {Scherrer},
  {Schwadron}, {Tapley}, \& {Tyler}}]{mccomas_etal:11a}
{McComas}, D.~J., {Carrico}, J.~P., {Hautamaki}, B., {et~al.} 2011, Space
  Weather, 9, 11002

\bibitem[{McComas {et~al.}(2017)McComas, Zirnstein, Bzowski, Dayeh, Funsten,
  Fuselier, Janzen, Kubiak, Kucharek, M{\"o}bius, Reisenfeld, Schwadron,
  Sok{\o}{\l}, Szalay, \& Tokumaru}]{mccomas_etal:17a}
McComas, D.~J., Zirnstein, E.~J., Bzowski, M., {et~al.} 2017, \apjs, 229, 41

\bibitem[{{M{\"o}bius} {et~al.}(2009){M{\"o}bius}, {Kucharek}, {Clark},
  {O'Neill}, {Petersen}, {Bzowski}, {Saul}, {Wurz}, {Fuselier}, {Izmodenov},
  {McComas}, {M{\"u}ller}, \& {Alexashov}}]{mobius_etal:09a}
{M{\"o}bius}, E., {Kucharek}, H., {Clark}, G., {et~al.} 2009, \ssr, 146, 149

\bibitem[{M{\"o}bius {et~al.}(2012)M{\"o}bius, Bochsler, Heirtzler, Kucharek,
  Lee, Leonard, Petersen, Schwadron, Valocvin, Wu, Bzowski, Kubiak, Fuselier,
  Saul, Wurz, McComas, \& Crew}]{mobius_etal:12a}
M{\"o}bius, E., Bochsler, P., Heirtzler, D., {et~al.} 2012, \apjs, 198, 11

\bibitem[{{M{\"o}bius} {et~al.}(2015{\natexlab{a}}){M{\"o}bius}, {Bzowski},
  {Fuselier}, {Heirtzler}, {Kubiak}, {Kucharek}, {Lee}, {Leonard}, {McComas},
  {Schwadron}, {Sok\'{o}{\l}}, \& {Wurz}}]{mobius_etal:15b}
{M{\"o}bius}, E., {Bzowski}, M., {Fuselier}, S.~A., {et~al.}
  2015{\natexlab{a}}, \apjs, 220, 24

\bibitem[{{M{\"o}bius} {et~al.}(2015{\natexlab{b}}){M{\"o}bius}, {Bzowski},
  {Fuselier}, {Heirtzler}, {Kubiak}, {Kucharek}, {Lee}, {Leonard}, {McComas},
  {Schwadron}, {Sok{\'o}{\l}}, \& {Wurz}}]{mobius_etal:15a}
---. 2015{\natexlab{b}}, Journal of Physics: Conference Series, 577, 012019

\bibitem[{Park {et~al.}(2016)Park, Kucharek, M{\"o}bius, Galli, Kubiak,
  Bzowski, \& McComas}]{park_etal:16a}
Park, J., Kucharek, H., M{\"o}bius, E., {et~al.} 2016, \apj, 833, 130

\bibitem[{Park {et~al.}(2015)Park, Kucharek, M{\"o}bius, Galli, Livadiotis,
  Fuselier, \& J.McComas}]{park_etal:15a}
---. 2015, \apjs, 220, 34

\bibitem[{Ruci{\'n}ski {et~al.}(2003)Ruci{\'n}ski, Bzowski, \&
  Fahr}]{rucinski_etal:03}
Ruci{\'n}ski, D., Bzowski, M., \& Fahr, H.~J. 2003, \ag, 21, 1315

\bibitem[{Saul {et~al.}(2012)Saul, Wurz, M{\"o}bius, Bzowski, Fuselier, Crew,
  Rodriguez, Leonard, McComas, Schwadron, Bochsler, \& Scheer}]{saul_etal:12a}
Saul, L., Wurz, P., M{\"o}bius, E., {et~al.} 2012, \apjs, 198, 14

\bibitem[{Saul {et~al.}(2013)Saul, Bzowski, Fuselier, Kubiak, McComas,
  M{\"o}bius, Sok{'o}{\l}, Rodr{\'i}guez, Scheer, \& Wurz}]{saul_etal:13a}
Saul, L., Bzowski, M., Fuselier, S., {et~al.} 2013, \apj, 767, 130

\bibitem[{{Schwadron} {et~al.}(2015){Schwadron}, {M{\"o}bius}, {Leonard},
  {Fuselier}, {Bzowski}, {Frisch}, {Heirtzler}, {Kubiak}, {Kucharek}, {Lee},
  {McComas}, {Rahmanifard}, {Sok\'{o}{\l}}, \& {Swaczyna}}]{schwadron_etal:15a}
{Schwadron}, N., {M{\"o}bius}, E., {Leonard}, T., {et~al.} 2015, \apjs, 220, 25

\bibitem[{{Schwadron} {et~al.}(2013){Schwadron}, {Moebius}, {Kucharek}, {Lee},
  {French}, {Saul}, {Wurz}, {Bzowski}, {Fuselier}, {Livadiotis}, {McComas},
  {Frisch}, {Gruntman}, \& {Mueller}}]{schwadron_etal:13a}
{Schwadron}, N.~A., {Moebius}, E., {Kucharek}, H., {et~al.} 2013, \apj, 775, 86

\bibitem[{{Schwadron} {et~al.}(2014){Schwadron}, {Moebius}, Fuselier, McComas,
  Funsten, Janzen, Reisenfeld, Kucharek, Lee, Fairchild, Allegrini, Dayeh,
  Livadiotis, Reno, Bzowski, Sok{\'o}{\l}, Kubiak, Christian, DeMajistre,
  Frisch, Galli, Wurz, \& Gruntman}]{schwadron_etal:14b}
{Schwadron}, N.~A., {Moebius}, E., Fuselier, S., {et~al.} 2014, \apjs, 215, 13

\bibitem[{{Sok{\'o}{\l}} \& {Bzowski}(2014)}]{sokol_bzowski:14a}
{Sok{\'o}{\l}}, J.~M., \& {Bzowski}, M. 2014, ArXiv e-prints, arXiv:1411.4826

\bibitem[{{Sok\'{o}{\l}} {et~al.}(2016){Sok\'{o}{\l}}, {Bzowski}, {Kubiak}, \&
  {M{\"o}bius}}]{sokol_etal:16a}
{Sok\'{o}{\l}}, J.~M., {Bzowski}, M., {Kubiak}, M., \& {M{\"o}bius}, E. 2016,
  \mnras, 458, 3691

\bibitem[{{Sok\'{o}{\l}} {et~al.}(2015{\natexlab{a}}){Sok\'{o}{\l}}, {Kubiak},
  {Bzowski}, \& {Swaczyna}}]{sokol_etal:15b}
{Sok\'{o}{\l}}, J.~M., {Kubiak}, M., {Bzowski}, M., \& {Swaczyna}, P.
  2015{\natexlab{a}}, \apjs, 220, 27

\bibitem[{{Sok\'{o}{\l}} {et~al.}(2015{\natexlab{b}}){Sok\'{o}{\l}}, {Bzowski},
  {Kubiak}, {Swaczyna}, {Galli}, {Wurz}, {M{\"o}bius}, {Kucharek}, {Fuselier},
  \& {McComas}}]{sokol_etal:15a}
{Sok\'{o}{\l}}, J.~M., {Bzowski}, M., {Kubiak}, M., {et~al.}
  2015{\natexlab{b}}, \apjs, 220, 29

\bibitem[{{Swaczyna} {et~al.}(2015){Swaczyna}, {Bzowski}, {Kubiak},
  {Sok\'{o}{\l}}, {M{\"o}bius}, {Leonard}, {Heirtzler}, {Kucharek},
  {Schwadron}, {Fuselier}, \& {McComas}}]{swaczyna_etal:15a}
{Swaczyna}, P., {Bzowski}, M., {Kubiak}, M., {et~al.} 2015, \apjs, 220, 26

\bibitem[{{Tarnopolski} \& {Bzowski}(2009)}]{tarnopolski_bzowski:09}
{Tarnopolski}, S., \& {Bzowski}, M. 2009, \aap, 493, 207

\bibitem[{Wieser \& Wurz(2005)}]{wieser_wurz:05a}
Wieser, M., \& Wurz, P. 2005, Measurement Science and Technology, 16, 2511

\bibitem[{{Witte}(2004)}]{witte:04}
{Witte}, M. 2004, \aap, 426, 835

\bibitem[{{Wood} {et~al.}(2015){Wood}, {M{\"u}ller}, \&
  {Witte}}]{wood_etal:15a}
{Wood}, B.~E., {M{\"u}ller}, H.-R., \& {Witte}, M. 2015, \apj, 801, 62

\bibitem[{Wurz {et~al.}(2008)Wurz, Saul, Scheer, M{\"o}bius, Kucharek, \&
  Fuselier}]{wurz_etal:08a}
Wurz, P., Saul, L., Scheer, J.~A., {et~al.} 2008, J. Appl. Phys., 103, 054904

\bibitem[{Wurz {et~al.}(2006)Wurz, Scheer, \& Wieser}]{wurz_etal:06a}
Wurz, P., Scheer, J., \& Wieser, M. 2006, {e-J. of Surface Science and
  Nanotechnology}, 4, 394

\bibitem[{{Zirnstein} {et~al.}(2016){Zirnstein}, {Heerikhuisen}, {Funsten},
  {Livadiotis}, {McComas}, \& {Pogorelov}}]{zirnstein_etal:16b}
{Zirnstein}, E.~J., {Heerikhuisen}, J., {Funsten}, H.~O., {et~al.} 2016, \apjl,
  818, L18

\end{thebibliography}

%% Include this line if you are using the \added, \replaced, \deleted
%% commands to see a summary list of all changes at the end of the article.
%\listofchanges

\end{document}